\documentclass[a4paper,11pt]{article}
\usepackage{jheppub}% for details on the use of the package, please see the JINST-author-manual
\usepackage{tcolorbox}
\newcommand{\e}{\epsilon}
\newcommand{\eeqref}[1]{Eq. \eqref{#1}}

\newcommand{\linefill}{% a variation on \rightarrowfill
  {-}\mkern-7mu
  \cleaders\hbox{$\mkern-2mu-\mkern-2mu$}\hfill
  \mkern-7mu{-}%
}

\title{\boldmath {\huge \texttt{MultiHypExp}}: A \textsc{Mathematica} Package For Expanding Multivariate Hypergeometric Functions In Terms Of Multiple Polylogarithms}

\author[a,b]{Souvik Bera}
\affiliation[a]{Centre for High Energy Physics, Indian Institute of Science, Bangalore-560012, Karnataka, India}
\affiliation[b]{Physik-Institut, Universit\"at Z\"urich, Winterthurerstrasse 190, CH-8057 Z\"urich, Switzerland}

\emailAdd{souvikbera@iisc.ac.in}

\abstract{We present the \textsc{Mathematica} package \texttt{MultiHypExp} that allows for the expansion of multivariate hypergeometric functions (MHFs), especially those likely to appear as solutions of multi-loop, multi-scale Feynman integrals, in the dimensional regularization parameter. The series expansion of MHFs can be carried out around integer values of parameters to express the series coefficients in terms of multiple polylogarithms. The package uses a modified version of the algorithm prescribed in  \cite{Bera:2022ecn}. In the present work, we  relate a given MHF to a Taylor series expandable MHF by a differential operator. The Taylor expansion of the latter MHF is found by first finding the associated  partial differential equations (PDEs) from its series representation. We then bring the PDEs to the Pfaffian system and further to the canonical form, and solve them order by order in the expansion parameter using appropriate boundary conditions. The  Taylor expansion so obtained and the differential operators are used to find the series expansion of the given MHF.  We provide examples to demonstrate the algorithm and  to describe the usage of the package, which  can be found in this \href{https://github.com/souvik5151/MultiHypExp}{url}.\\

\vspace{1cm}
\textbf{Program summary}\\
\textit{Program Title:} \texttt{MultiHypExp.wl} version 1.0\\
\textit{Developer’s repository link:} \url{https://github.com/souvik5151/MultiHypExp}\\
\textit{Licensing provisions:} GNU General Public License 3\\
\textit{Programming language:} Wolfram Mathematica version 11.3 or higher\\
\textit{External routines/libraries:} HolonomicFunctions, CANONICA, HYPERDIRE, PolyLogTools, HPL\\
\textit{Nature of problem:} Obtaining the series expansion of multivariate hypergeometric functions around integer values of Pochhammer parameters in terms of multiple polylogaritms\\
\textit{Solution method:} Mathematica implementation of a modified version of the algorithm prescribed in Nucl. Phys. B 989 (2023) 116145.
}

\begin{document}
\maketitle
\flushbottom

\section{Introduction }
\label{sec:intro}

%\cite{Gelfand1992GeneralHS,Gelfand1994DiscriminantsRA,Im1990GeneralizedEI,delaCruz:2019skx,Klausen:2019hrg,Feng:2022kgh,Klausen:2021yrt,Pal:2021llg,Feng:2019bdx,Tellander:2021xdz,Ananthanarayan:2022ntm,Zhang:2023fil,Chestnov:2022alh,Munch:2022ouq,Chestnov:2023kww}.

Multivariate Hypergeometric Functions \cite{Bateman:1953,Bailey:1964,Slater:1966,Exton:1976,Srivastava:1985,aomoto2011theory} (hereinafter MHFs) while being ubiquitous in mathematics and physics, are also of great importance today to elementary particle physics applications as they appear as solutions to the dimensionally regularized multi-scale Feynman integrals required for higher-order corrections to the scattering amplitude. The expansion of Feynman integrals, expressed in terms of MHFs, in the dimension parameter $\e = (4-D)/2$ necessitates the construction of robust algorithms and efficient computer programs. The purpose of the present work is to develop the considerations of the recent publication \cite{Bera:2022ecn} further and present a \textsc{Mathematica} \cite{Mathematicav11.3} realization in terms of the \texttt{MultiHypExp} package.

A long-standing relationship between the Feynman integrals and MHFs has existed since the conjecture of Regge \cite{Regge}. Standard methods like Mellin-Barnes \cite{Dubovyk:2022obc,AnantPRL2021}, Negative DIMension \cite{Halliday:1987an,Gonzalez:2007ry}, Method Of Brackets \cite{GONZALEZ201050,gonzalez2010method,Ananthanarayan:2021not}, Functional Equations \cite{Tarasov:2022clb} can yield hypergeometric function (HF) representation of Feynman integrals.  The connection between these two is further revived by recent works on realizing Feynman integrals as GKZ hypergeometric systems 
\cite{Gelfand1992GeneralHS,Gelfand1994DiscriminantsRA,Im1990GeneralizedEI,delaCruz:2019skx,Klausen:2019hrg,Ananthanarayan:2022ntm}.
In the dimensional regularization, the dimension $D$ of the  Feynman integrals appears linearly as Pochhammer parameters of its associated HF representations, which  are further expanded in series in the parameter $\e$.  Series expansion of one variable HFs are well studied in the literature \cite{Kalmykov:2006pu,Kalmykov:2007pf,Kalmykov:2008ge,Greynat:genhypfun}.  The $\e$-expansion of double variable Appell and Kamp\'e de F\'eriet are discussed in \cite{Greynat:appellkdf,Greynat:2014jsa,Moch:2001zr,Weinzierl:halfint}. Certain MHFs related to Feynman integrals are expanded in series using differential equation method \cite{Yost:2011wk,Bytev:2012ud,Kalmykov:2020cqz}. The series expansion of multi-integrals or multi-sums over hyperexponential and/or hypergeometric functions are considered in \cite{Bierenbaum:2007zu,Blumlein:2010zv,Blumlein:2011kef,Ablinger:2012ph,Schneider:2013zna,Blumlein:2021hbq}. Automated packages also exist to find the series coefficients of certain HFs analytically \cite{HypExp,HypExp2,Moch:Xsummer,Weinzierl:code,Ablinger:2013cf} and numerically \cite{NumExp}.

Recently, a method to expand any MHF in series around its Pochhammer parameter has been prescribed  in  \cite{Bera:2022ecn}. In this approach, starting with the series representation of a given MHF, the expansion can be obtained around any value of its parameters, and the series coefficients are expressed in terms of MHFs having the same domain of convergences. However, it is possible to express the coefficients in terms of multiple polylogarithms (MPLs) if the series expansion is carried out around integer values of the Pochhammer parameters for most of the MHFs (see the Section \ref{sec:mathematica} below for the exceptional cases). It is advantageous to express the series expansion coefficients in terms of MPLs, whenever possible, rather than MHFs, as the former can be readily evaluated at any value of their arguments using well-established computer libraries, even though the series representation of the given MHF is valid in certain domains of convergence. 

In this work, we reshape the algorithm presented in \cite{Bera:2022ecn} such that, whenever possible, the series coefficients of MHFs can be expressed in terms of much simpler MPLs rather than MHFs. The modified algorithm is further implemented in \textsc{Mathematica} as the \texttt{MultiHypExp} package. In its present version, it can obtain a series expansion of one variable $_pF_{p-1}$, double variable Appell-Horn, certain Kamp\'e de F\'eriet functions and certain triple variable Lauricella-Saran functions around integer values their Pochhammer parameters, which typically arise in the Feynman integral calculus.

 The article is organized as follows. We recapitulate the algorithm proposed in \cite{Bera:2022ecn} and discuss the modifications made in Section \ref{sec:method}, followed by illustrative examples of Gauss $_2F_1$ and Appell $F_2$ functions in Section \ref{sec:examples}.   An application to Feynman integral is provided in Section \ref{sec:applications}.  We give a detailed description of the commands of  the package and their usage in Section \ref{sec:demo} and explain how the methodology is implemented in \textsc{Mathematica} in Section \ref{sec:mathematica}. Finally, we present our conclusions in Section \ref{sec:Conclusions} which is followed by number of appendices. In Appendix \ref{sec:definitionMPL} and \ref{sec:definitionMHFs} we provide the definitions of MPLs and MHFs respectively. A list of expressions of series expansions of pure MHFs is given in Appendix \ref{sec:exactresult}. In Appendix \ref{sec:Transformation_Theory} and \ref{sec:Reduction_formulae}, we discuss the transformation theory and reduction formulae of MHFs.

\section{Methodology}
\label{sec:method}

A method to obtain series expansion of any MHF around arbitrary values of Pochhammer parameters is prescribed in \cite{Bera:2022ecn}, where the expansion parameter, say $\e$, appears linearly in the Pochhammer parameters.  Throughout this method, the hypergeometric structure of a given MHF remains intact. In other words, starting with a series presentation, one can find the series expansion of a given MHF, whose  coefficients are expressed in terms of higher summation fold MHFs with the same domain of convergence as that of the given MHF. However, the higher fold MHFs appearing in the series coefficients contain the same number of independent variables as the given MHF. For instance, the one variable Gauss $_2F_1(\e,-\e;\e-1;x)$  is expanded in series in \cite{Bera:2022ecn} as
\begin{align} \label{eqn:2f1exp}
    _2F_1(\e,-\e;\e-1;x)&= 1 - \e^2 x~ \, _3F_2(1,1,1;2,2;x) 
+ \e^3 \Bigg(\frac{2}{3} x \, _3F_2(1,1,1;2,2;x) \nonumber\\
&+\frac{x}{3}~ \text{KdF}^{2:1;2}_{2:0;1}
	\left[
	\setlength{\arraycolsep}{0pt}% local assignment
	\begin{array}{c@{{}:{}}c@{;{}}c}1,1&1&1,1
		\\[1ex]
		2,2&\linefill&2
	\end{array}
	\;\middle|\;
	x,x
	\right]+ \frac{x^2}{12}~ \text{KdF}^{2:1;2}_{2:0;1}
	\left[
	\setlength{\arraycolsep}{0pt}% local assignment
	\begin{array}{c@{{}:{}}c@{;{}}c}2,2&1&1,2
		\\[1ex]
		3,3&\linefill&3
	\end{array}
	\;\middle|\;
	x,x
	\right]
\Bigg) \nonumber\\
&+ O(\e^4)
\end{align}
Here, the Kamp\'e de F\'eriet functions (denoted as KdF) are double summation fold HFs with only one variable $x$.

The same Gauss $_2F_1(\e,-\e;\e-1;x)$ can be expanded in series about its argument as
\begin{align*}
    _2F_1(\e,-\e;\e-1;x) = 1-\frac{x \epsilon ^2}{\epsilon -1}+\frac{1}{2} x^2 \epsilon  (\epsilon +1)-\frac{1}{6} x^3 ((\epsilon -2) \epsilon  (\epsilon +2))+O\left(x^4\right)
\end{align*}
In this article, we focus on the series expansion of MHFs about its Pochhammer parameters as in \eeqref{eqn:2f1exp} and not the series expansion about its arguments. Throughout the article,  unless otherwise indicated, the term `series expansion' or `series coefficients' of MHFs means the series expansion or the coefficients of series expansion of MHFs about its Pochhammer parameters. 

It is well-known that the function $_3F_2(1,1,1;2,2;x) $ can be written in terms of simpler function 
\begin{align}
    _3F_2(1,1,1;2,2;x)  = \frac{\text{Li}_2(x)}{x}
\end{align}
Nonetheless, such a reduction formulae for the KdF functions in \eeqref{eqn:2f1exp} is difficult to obtain. To the best of our knowledge, there do not exist reduction formulae for arbitrary MHFs.

There are some advantages in expressing the coefficients of the series expansion of MHFs (whenever possible) in terms of simpler functions such as  MPLs \cite{goncharov2011multiple,goncharov2001multiple}, whose properties are well studied. The numerical evaluation of the multiple polylogs for any values of its arguments can be easily performed using readily available  computer libraries \cite{Gehrmann:2001pz,Maitre:2005uu,Maitre:2007kp,Buehler:2011ev,Vollinga:2004sn,Ablinger:2018sat,FastGPL,handyG}, whereas the series representation of the MHFs are valid in certain domain of convergences. For instance, the KdF functions  in \eeqref{eqn:2f1exp} are only valid for $|x|<1$.
This motivates us to modify the algorithm  presented in  \cite{Bera:2022ecn} slightly such that, the series coefficients of a given MHF, whenever possible, can be expressed in terms of MPLs. In order to achieve this, we must perform the series expansion around integer values of Pochhammer parameters for most of the HFs, because the series expansion around rational values of parameters often leads to functions beyond MPLs.

We now recapitulate the algorithm proposed in \cite{Bera:2022ecn} below, which consists of five steps.
\begin{itemize}
    \item \textit{Step 1} : Distinguish the type of series expansion by observing the Pochhammer parameters of the given MHF (say $F(\e)$)
    \item \textit{Step 2} : Find the series expansion of $F(\e)$, if it is of Taylor type
    \item \textit{Step 3} : If the series expansion is of Laurent type, find a secondary function, say $G(\e)$ that can be related to $F(\e)$ by a differential operator $H(\e)$
    \begin{align}
        F(\e) = H(\e) \bullet G(\e)
    \end{align}
    and $G(\e)$ can be expanded in Taylor series following \textit{Step 2}. Here, the symbol $\bullet$ means the action of the differential operator $H(\e)$ on the function $G(\e)$.
    \item \textit{Step 4} : Find the corresponding differential operator $H(\e)$
    \item \textit{Step 5} : Perform the series expansion of the operator $H(\e)$ and apply it on the Taylor expansion of $G(\e)$ and collect different powers of $\e$ 
\end{itemize}

We discuss each of the steps in detail.

\subsection{\textit{Step 1} : Determination of the type of the series expansion}
\label{sec:step1}

By observing the structure of the Pochhammer parameters in the series representation of a given MHF, one may find the type of the series expansion of the same. The series expansion of a MHF may be of Laurent series if any of the following situations appear
\begin{enumerate}
    \item When one or more lower Pochhammer parameters (i.e., Pochhammer parameters in the denominator) are of the form 
    \begin{align*}
        (n + q \e)_p
    \end{align*}
    \item When one or more upper Pochhammer parameters (i.e., Pochhammer parameters in the numerator) are of the form 
    \begin{align*}
        (p + q \e)_n
    \end{align*}
    
\end{enumerate}
where $n$ is non positive integer and $p$ is non negative integer. We call a Pochhammer parameter \textit{singular} if it satisfies any of the above two conditions.   For instance, the series expansion of $_2F_1 (1,1;-1+\e;x)$ is of Laurent type because of the presence of the lower Pochhammer parameter $(-1+\e)_p$, where $p$ is the summation index running over non-negative integer values.

However, the above two criteria do not guarantee the fact that, the series expansion of the corresponding MHF must be of Laurent type. The expansion of $_2F_1 (\e,-\e;-1+\e;x)$ (given in \eeqref{eqn:2f1exp}) is of Taylor type even though the first condition is satisfied. This is due to the presence of  $\e$ dependent upper Pochhammer parameters. Thus, the above two criteria are necessary but not sufficient conditions.

\subsection{\textit{Step 2} : Taylor expansion of MHF}

In \cite{Bera:2022ecn}, the Taylor expansion of a MHF is found by taking successive derivatives of the function with respect to its Pochhammer parameters, which results in higher summation fold MHFs in the series coefficients. In order to express the series coefficients in terms of MPLs, we modify this step. 

To proceed, we define a HF with $n$ variables having Pochhammer parameters $\textbf{a}$ and $\textbf{b}$
\begin{align} \label{eqn:mhfdefinition}
		\textbf{F}:=F(\textbf{a};\textbf{b};\textbf{x}) = \sum_{\textbf{m}\in \mathbb{N}_0^r}  \frac{\Gamma(\textbf{a}+ \mu . \textbf{m} )/ \Gamma(\textbf{a}) }{\Gamma(\textbf{b}+ \nu . \textbf{m} )/ \Gamma(\textbf{b}) } \frac{\textbf{x}^\textbf{m}}{\textbf{m}!} = \sum_{\textbf{m}\in \mathbb{N}_0^r}  \frac{(\textbf{a})_{\mu . \textbf{m}} }{(\textbf{b})_{\nu . \textbf{m}} } \frac{\textbf{x}^\textbf{m}}{\textbf{m}!} = \sum_{\textbf{m}\in \mathbb{N}_0^r} A(\textbf{m}) \textbf{x}^\textbf{m}
	\end{align}
We have used vector notation here; $\textbf{a}, \textbf{b}$ and $\textbf{m}$ are vectors of length $p,q$ and $n$ respectively. $\mu$ and $\nu$ are matrices of size $p \times n$ and $q \times n$ respectively with integers as their elements. $\mathbb{N}_0$ denotes natural numbers including zero.
	\begin{itemize}
		\item $\textbf{a} := \{a_1 , a_2, \dots, a_n\}$
		\item $\textbf{x}^\textbf{m} := \prod_{i=1}^n x_i^{m_i}$
		\item $\textbf{m}! := \prod_{i=1}^n (m_i!)$
		\item $\Gamma(\textbf{a}) := \prod_{i=1}^n \Gamma(a_i)$
		\item $(\textbf{a})_\textbf{m} :=\prod_{i=1}^n (a_i)_{m_i} =\prod_{i=1}^n \Gamma(a_i+m_i)/\Gamma(a_i) $
		\item $\sum_{\textbf{m}\in \mathbb{N}_0^n} := \sum_{m_1 =0}^\infty \dots \sum_{m_n =0}^\infty$
	\end{itemize}
	
	These MHFs are known to satisfy partial differential equations \cite{Bateman:1953}. Let,
	\begin{align} \label{eqn:annihilator1}
		P_i = \frac{A(\textbf{m}+\textbf{e}_\textbf{i})}{A(\textbf{m})} = \frac{g_i(\textbf{m})}{h_i(\textbf{m})} ,\hspace{1cm}\text{~} i=1,\dots,n
	\end{align}
	where $\textbf{e}_\textbf{i}$ is unit vector with $1$ in its i-th entry. The annihilators $L_i$ of  $	F(\textbf{a};\textbf{b};\textbf{x})$ (\eeqref{eqn:mhfdefinition}) are given by
	\begin{align} \label{eqn:annihilator2}
		L_i = \left[ h_i(\pmb{\theta}) \frac{1}{x_i} - g_i(\pmb{\theta}) \right]
	\end{align}
	where $\pmb{\theta} = \{\theta_1,\dots,\theta_n \}$ is a vector containing  Euler operators $\theta_i= x_i\partial_{x_i}$.

    The set of PDEs associated with a MHF can be brought to Pfaffian form 
    \begin{align} \label{eqn:Pfaffian}
        d \textbf{g} = \Omega \textbf{g}
    \end{align}
    where $\Omega =  \sum_{i=1}^n \Omega_i dx_i $ and the vector $\textbf{g}$ contains the function $\textbf{F}$ and its derivatives. 

    \begin{align}
        \textbf{g} = (\textbf{F}~~,~~ \pmb{\theta_i}\bullet \textbf{F}~~,~~ \pmb{\theta_i \theta_j}\bullet \textbf{F}, \dots )^T
    \end{align}
    
    The integrability condition reads $d \Omega + \Omega \wedge \Omega =0$. The length of $\textbf{g}$ is equal to the holonomic rank of the system of PDEs, which can be computed using the Gr\"obner basis calculation. The  Pfaffian system of Appell and Lauricella functions are well studied in mathematics literature \cite{yoshida2013fuchsian,Kato_Pfaff_F4,Matsumoto_Pfaff_F4,Matsumoto_Pfaff_FA,Matsumoto_Pfaff_FD}.

    We find the Taylor series expansion of $\textbf{F}$ by finding the solution of \eeqref{eqn:Pfaffian} with appropriate boundary condition, to be described below. In the context of computing Feynman integrals using the differential equation method, it is observed in \cite{Henn:2013pwa,Henn:2014qga} that, by choosing a particular set of master integrals, the set of differential equations can be brought to a much simpler form known as the \textit{canonical form}. We apply this technique to solve the system \eeqref{eqn:Pfaffian}.
    
    Starting with the Pfaffian system
    \begin{align} 
         d \textbf{g} = \Omega(\e) \textbf{g}
    \end{align}
    
    One can find a transformation $T$ to bring the system into canonical form
        \begin{align}\label{eqn:canonicalform}
            d \textbf{g}' = \e \Omega' \textbf{g}'
        \end{align}
        with 
        \begin{align}
            \textbf{g} &=  T  \label{eqn:transformation}\textbf{g}' \\ \Omega' &=  T^{-1}\Omega T - T^{-1} dT       
\end{align}
The system \eeqref{eqn:canonicalform} can be solved order by order in $\e$ with a boundary condition.

Algorithms exist for calculating the transformation matrix that converts the Pfaffian system to the canonical form \cite{Moser,Meyer:2016slj, Lee:2014ioa, Argeri:2014qva}. Additionally, there are readily accessible computer programs that have been developed to implement these algorithms. \cite{Meyer:2017joq, Prausa:2017ltv, Gituliar:2017vzm,Lee:2020zfb}

For any MHF, we note that
\begin{align*}
&\left.F(\textbf{a};\textbf{b};\textbf{x})\right|_{\textbf{x} = \textbf{0}}  = 1\\
\pmb{\theta_i}\bullet &\left.F(\textbf{a};\textbf{b};\textbf{x})\right|_{\textbf{x} = \textbf{0}}  = 0\\
\pmb{\theta_i\theta_j}\bullet &\left.F(\textbf{a};\textbf{b};\textbf{x})\right|_{\textbf{x} = \textbf{0}}  = 0\\
&\vdots
\end{align*}

Therefore, the boundary condition can be easily obtained in terms of $\textbf{g}$ as,
\begin{align}
    \left. \textbf{g}\right|_{\textbf{x}=\textbf{0}} = (1,0,0,\dots)^T
\end{align}

The system \eeqref{eqn:canonicalform} can now be solved with the boundary condition
\begin{align}\label{eqn:BCg'}
   \left. \textbf{g}'\right|_{\textbf{x}=\textbf{0}} = T^{-1} \left. \textbf{g}\right|_{\textbf{x}=\textbf{0}}
\end{align}

Finally, the series expansion of $\textbf{F}$ can be found by converting the solution of \eeqref{eqn:canonicalform} with boundary condition \eeqref{eqn:BCg'} to the $\textbf{g}$ basis by using \eeqref{eqn:transformation}.

\subsection{\textit{Step 3} : Construction of the secondary function}
\label{sec:step3}

The secondary function $G(\e)$ related to $F(\e)$ can be obtained by performing the following replacements of the Pochhammer parameters of $F(\e)$
\begin{enumerate}
    \item When one or more lower Pochhammer parameters of $F(\e)$ are singular, then
    \begin{align}
        (n + q \e)_p &\rightarrow (1+ q\e)_p
    \end{align}
     \item When one or more upper Pochhammer parameters of $F(\e)$ are singular, then
    \begin{align}
        (p + q \e)_n &\rightarrow (q\e)_n
    \end{align}
\end{enumerate}

As an example, the function $~_2F_1 (\e,-\e;-1+\e;x)$ contains a lower Pochhammer parameter that is singular. Therefore, following the replacement rule 1 above, we replace the lower Pochhammer parameter to find the associated secondary function as
\begin{align}
    _2F_1 (\e,-\e;1+\e;x)
\end{align}

Thus obtained secondary function $G(\e)$ can be expanded in the Taylor series following \textit{Step 2}.

\subsection{\textit{Step 4} : The differential operator}

There exist differential operators that relate two MHFs with Pochhammer parameters differed by integer values. In \cite{Takayama1989}, a general algorithm based on Gr\"obner basis techniques is provided by Takayama. These differential operators play crucial role in the differential reduction of MHFs \cite{Hyperdire,Hyperdire2,Hyperdire3,Hyperdire4}. For our purpose, we only need two types of differential operators; the step-down operator for the lower Pochhammer parameters and step-up operator for the upper Pochhammer parameters. These two types of operators can be easily obtained from the series representation of a MHF.  For a MHF with $n$ variables (\eeqref{eqn:mhfdefinition}), the step-down operators for the lower Pochhammer parameters are given by

\begin{align}\label{eqn:stepdownop}
	F(\textbf{a};\textbf{b};\textbf{x}) = 	\frac{1}{b_i}\left(\sum_{j=1}^{n} \nu_{i j} \theta_{x_j} +b_i\right) \bullet	F(\textbf{a};\textbf{b}+\mathbf{e_i};\textbf{x}) = H_-(b_i) \bullet  F(\textbf{a};\textbf{b}+\mathbf{e_i};\textbf{x})
	\end{align}

 and the step-up operators for the upper Pochhammer parameters can be found as
 \begin{align}\label{eqn:stepupop}
F(\textbf{a},\textbf{b},\textbf{x}) = 	\frac{1}{a_i-1}\left(\sum_{j=1}^{n} \mu_{i j} \theta_{x_j} +a_i-1\right) \bullet	F(\textbf{a}-\mathbf{e_i},\textbf{b},\textbf{x}) = H_+(a_i) \bullet  F(\textbf{a}-\mathbf{e_i},\textbf{b},\textbf{x})
 \end{align}
	
	Here, as before, the bullet means the action of the operator $H_\pm$ on a function. When required, the step-up and the step-down operators may be applied multiple times to increase or decrease any upper or lower Pochhammer parameter by  suitable integer. In such cases, the product of the unit operators modulo the ideal generated by $L_i$'s can be taken as the required differential operator.
	
	\begin{align}
		H = \left[\prod_{i,j} H_-(b_i) H_+(a_j) \right]/ \langle L_1,\dots,L_n\rangle
	\end{align}

\subsection{\textit{Step 5} : Action of the differential operator}

In the final step, we apply the differential operator found in \textit{Step 4} on the Taylor expansion of $G(\e)$ obtained in \textit{Step 3}. Since the coefficients of the Taylor expansion of $G (\e)$ are expressed in terms of MPLs, the task of performing the application of the differential operator is the same as finding the ordinary derivatives of the MPLs with respect to their arguments.

Next, we present examples of series expansion of one and two-variable HFs to illustrate the methodology.

\section{Examples}
\label{sec:examples}

\subsection{Gauss $_2 F_1$ function}
\label{sec:examples2f1}
Let us consider the following Gauss HF, whose series expansion we wish to find.
\begin{align}\label{eqn:examples2f1def}
     F(\e) := ~_2F_1(\e,-\e;\e-1;x) =  \sum_{m=0}^\infty \frac{ (\e)_m (-\e)_m}{ (\e-1)_m} \frac{x^m}{m!}  
\end{align}

We notice that the lower Pochhammer parameter of $F(\e)$ is singular. Therefore, the series expansion may be of Laurent type. Therefore, we construct the secondary function.
The secondary function $G(\e)$, which is related to $F(\e)$ and has a Taylor series expansion, can be found by replacing the singular Pochhammer parameter,
\begin{align}
    G(\e) := ~_2F_1(\e,-\e;\e+1;x) =  \sum_{m=0}^\infty \frac{ (\e)_m (-\e)_m}{ (\e+1)_m} \frac{x^m}{m!} 
\end{align}

Next, we go on to find the series expansion of $G(\e)$. By constructing a vector $\textbf{g} = ( G(\e), \theta_x G(\e))^T$ and making good use of the ordinary differential equation of Gauss $_2F_1$, we obtain the following Pfaffian system 
\begin{align}
    d \textbf{g}  =  \Omega \textbf{g} \hspace{.5cm},\hspace{.5cm} \Omega = \left(
\begin{array}{cc}
 0 & \frac{1}{x} \\
 \frac{\epsilon ^2}{x-1} & \frac{\epsilon }{x-1}-\frac{\epsilon }{x} \\
\end{array}
\right)
\end{align}

Further, the Pfaffian system can be brought to canonical form by the transformation matrix, which can obtained using \texttt{CANONICA} \cite{Meyer:2017joq} 
\begin{align}
    T = \left(
\begin{array}{cc}
 1 & 0 \\
 0 & \epsilon  \\
\end{array}
\right)
\end{align}
which reads
\begin{align}
    d \textbf{g}'  =  \e \Omega' \textbf{g}' \hspace{.5cm},\hspace{.5cm}
    \Omega' = \left(
\begin{array}{cc}
 0 & \frac{1}{x} \\
 \frac{1}{x-1} & \frac{1}{x-1}-\frac{1}{x} \\
\end{array}
\right)
\end{align}

This system can now be solved order by order in $\e$ with the boundary condition given by
\begin{align}
    \textbf{g}(x=0) = (1,0)^T
\end{align}

Thus, we find the Taylor series expansion of $G(\e)$ as
\begin{align}
    G(\e) = 1+\epsilon ^2 G(0,1;x)+\epsilon ^3 (-G(0, 0, 1; x) + G(0, 1, 1;x))+O\left(\epsilon ^4\right)
\end{align}

Here $G(0,1;x)$'s are MPLs (see Appendix \ref{sec:definitionMPL} for their definition), not to be confused with the secondary function $G(\e)$.
Next, we obtain the differential operator that relates the two Gauss HFs
\begin{align}
    F(\e) =  H(\e) \bullet  G(\e)
\end{align}

Following the prescription given in \textit{Step 4}, we find
\begin{align}
    H (\e) = \frac{ (\e (2 x-1)-x+1)}{(\e-1) \e (x-1)} \theta_x +\frac{\e (2 x-1)-x+1}{(\e-1) (x-1)}
\end{align}

Finally, we apply, the differential operator $H(\e)$ on the Taylor expansion of $G(\e)$ to find the series expansion of $F(\e)$
\begin{align}
    F(\e) &= 1 + \e \Bigg[G(1;x)-\frac{x}{x-1}\Bigg] + \e^2 \Bigg[-\frac{x }{x-1}G(1;x)+G(1,1;x)-\frac{x}{x-1}\Bigg] + O(\e^3)
\end{align}

The result is consistent with the result obtained using the \texttt{HypExp} \cite{HypExp} package.

\subsection{Appell $F_2$ function}\label{sec:examplesF2}
Let us now consider an example of double variable Appell $F_2$ function. 
\begin{align}\label{eqn:examplesF2def1}
    F := F_2 (1,1,\e;\e,-\e;x,y) 
\end{align}

We observe that, both the lower Pochhammer parameters of the above function are singular. Thus the series expansion of $F_2$ may be of Laurent type.

Thus, following \textit{step 3}, we find a secondary function by replacing the singular lower Pochhammer parameters by non-singular ones
\begin{align}\label{eqn:examplesF2def2}
  F' = F_2 (1,1,\e;1+\e,1-\e;x,y) 
\end{align}
The secondary function $F'$ can  now be expanded in the Taylor series. Note that, the series expansion of this particular Appell $F_2$ function (i.e., $F'$) is readily available in the literature ( Eq. (81)  of \cite{Moch:2001zr}  and Section 3.5 of \cite{Bera:2022ecn} ).

Following \textit{step 2}, we find the series expansion of $F'$. To proceed, we obtain the PDE associated with the Appell $F_2(a, b_1, b_2; c_1, c_2 ;x,y)$ function
\begin{align}
    L_1 &=  -a b_1+\left(c_1-x \left(a+b_1+1\right)\right)\partial_x-b_1 y \partial_y-x y \partial_x \partial_y-(x-1) x \partial_x^2 \\
    L_2 &=  -a b_2+\left(c_2-y \left(a+b_2+1\right)\right)\partial_y -b_2 x \partial_x-x y \partial_x \partial_y-(y-1) y \partial_y^2
\end{align}
The operators $L_1$ and $L_2$ annihilate the Appell $F_2$ function. 

Next, using the vector 
\begin{align}
    \textbf{g} =  \left( F', \theta_x F', \theta_y F', \theta_x \theta_y F'  \right)^T
\end{align}
the above set of PDEs is brought to the Pfaff system
\begin{align}
    d \textbf{g}  =  [ \Omega_1 dx + \Omega_2 dy ] \textbf{g}
\end{align}
where,
\begin{small}
\begin{align}
\Omega_1 &=   \left(
\begin{array}{cccc}
 0 & \frac{1}{x} & 0 & 0 \\
 -\frac{1}{x-1} & \frac{\epsilon -2}{x-1}-\frac{\epsilon }{x} & -\frac{1}{x-1} & -\frac{1}{x-1} \\
 0 & 0 & 0 & \frac{1}{x} \\
 \frac{\epsilon }{x-1}-\frac{\epsilon }{x+y-1} & \frac{(2-\epsilon ) \epsilon }{x-1}-\frac{(2-\epsilon ) \epsilon }{x+y-1} & \frac{\epsilon }{x-1}-\frac{2 \epsilon +1}{x+y-1} & \frac{-\epsilon -2}{x+y-1}-\frac{-\epsilon  x+(\epsilon +1) x-x-\epsilon }{(x-1) x} \\
\end{array}
\right)\\
\Omega_2 &= \left(
\begin{array}{cccc}
 0 & 0 & \frac{1}{y} & 0 \\
 0 & 0 & 0 & \frac{1}{y} \\
 -\frac{\epsilon }{y-1} & -\frac{\epsilon }{y-1} & \frac{\epsilon }{y}+\frac{-2 \epsilon -1}{y-1} & -\frac{1}{y-1} \\
 \frac{\epsilon }{y-1}-\frac{\epsilon }{x+y-1} & \frac{\epsilon }{y-1}+\frac{\epsilon  (\epsilon +1)-3 \epsilon }{x+y-1} & \frac{2 \epsilon +1}{y-1}-\frac{2 \epsilon +1}{x+y-1} & \frac{-\epsilon -2}{x+y-1}+\frac{\epsilon }{y}+\frac{1}{y-1} \\
\end{array}
\right)
\end{align}
\end{small}

Here $\theta_x = x \partial_x$ and $\theta_y = y \partial_y$ are Euler operators. Further, the Pfaff system is brought to the canonical form by the transformation matrix $T$
\begin{align}
    T = \left(
\begin{array}{cccc}
 \frac{1}{x-1} & 0 & 0 & 0 \\
 \frac{2 \epsilon -x}{(x-1)^2} & \frac{\epsilon }{x-1} & -\frac{x \epsilon }{(x-1) (x+y-1)} & 0 \\
 -\frac{\epsilon }{x-1} & 0 & \frac{\epsilon }{x+y-1} & 0 \\
 \frac{\epsilon  \left(x^2+(y-1) x-2 y \epsilon \right)}{(x-1)^2 (x+y-1)} & -\frac{\epsilon ^2}{x-1} & \frac{x \epsilon  (-x+y \epsilon +1)}{(x-1) (x+y-1)^2} & \frac{\epsilon ^2}{x+y-1} \\
\end{array}
\right)
\end{align}

Finally, the system is solved order by order in $\epsilon$ with the boundary condition 
\begin{align}
    \textbf{g} (x = 0, y=0) =  (1,0,0,0)^T
\end{align}
which is valid for all orders in $\e$. Thus we find the series expansion of $F'$ to be 
\begin{align} \label{eqn:Fpexpansion}
    F' &= -\frac{1}{x-1}+ \e \Bigg[\frac{(G(1-y;x)-2 G(1;x)+G(1;y))}{x-1} \Bigg]+ \e^2 \frac{1}{x-1}\Bigg[2 G(1;x) G(1;y)\nonumber\\
    &-2 G(1;y) G(1-y;x)-G(0,1-y;x)+2 G(1,1-y;x)+2 G(1-y,1;x)
    \nonumber\\
    &-G(1-y,1-y;x)+2 G(0,1;x)-4 G(1,1;x)+G(0,1;y)-2 G(1,1;y)\Bigg] + O(\e^3)
\end{align}

Next in \textit{step 4}, we find the differential operator that relates $F$ with $F'$
\begin{align}
    F = H\bullet F'
\end{align}
Using the package \texttt{HYPERDIRE} \cite{Hyperdire} or by constructing the step-down operators, we obtain
\begin{align}
    H = 1 + \frac{1}{\epsilon } \theta_x -\frac{1}{\epsilon } \theta_y -\frac{1}{\epsilon ^2} \theta_x \theta_y
\end{align}

In the final \textit{step 5}, we perform the action of the operator $H$ on the series expansion of $F'$ to find the Laurent series expansion of $F$,
\begin{align} \label{eqn:Fexpansion}
    F &= \frac{x }{\e} \left(\frac{2}{(x-1)^2}-\frac{1}{(x+y-1)^2}\right) + \e^0\Bigg[ x \left(\frac{4}{(x-1)^2}-\frac{2}{(x+y-1)^2}\right) G(1;x)\nonumber\\
    &+2 x \left(\frac{1}{(x+y-1)^2}-\frac{1}{(x-1)^2}\right) G(1;y)+x \left(\frac{1}{(x+y-1)^2}-\frac{2}{(x-1)^2}\right) G(1-y;x)\nonumber\\
    &-\frac{(x+2 y-1) (x (2 x+3 y-3)-y+1)}{(x-1)^2 (x+y-1)^2}\Bigg] + O(\e)
\end{align}

We crosschecked the above result numerically and it is found to be consistent.

\section{Application to the Feynman integrals }
\label{sec:applications}

In \cite{Anastasiou:1999ui}, the one loop three-point function with two massive propagators and one off-shell leg is expressed in terms of double variable HFs using the NDIM method. The authors presented several HF representations  (Eqs. (3.23) - (3.27) and Eq. (3.18) of the abovementioned article) for the same Feynman integral, which are related by analytic continuations. Furthermore, starting with a HF representation valid in a certain kinematic region (namely, region IIb ) and using the integral representations of MHFs, the authors found the series expansion of the Feynman integral in Eq. (3.34) of \cite{Anastasiou:1999ui}. 

In principle, any analytic continuation of the Feynman integral can be used to find the series expansion. In this Section, we start with the HF representation of the Feynman integral valid in region IIIb, which contains Kamp\'e de F\'eriet  and Horn function, and find the series expansion using the  \texttt{MultiHypExp} package.  By doing so, besides validating their results, we also examine if series expansions from different analytic continuations agrees or not.

%the task of performing the  series expansion of a MHF first and then obtaining the analytic continuation of series coefficients is equivalent to performing an analytic continuation of the MHF and then finding the series expansion of the resulting analytic continuation.

In region IIIb, the Feynman integral can be expressed as
\begin{align}
   I_3^D\left(\nu_1, \nu_2, \nu_3 ; Q_1^2, 0,0, M_1^2, M_2^2, 0\right)=I_3^{\left\{m_2, q_1\right\}}+I_3^{\left\{p_1, q_1\right\}}
\end{align}
where
\begin{align}
& I_3^{\left\{m_2, q_1\right\}}=(-1)^{\frac{D}{2}}\left(-M_1^2\right)^{\frac{D}{2}-\nu_1-\nu_2-\nu_3} \frac{\Gamma\left(\nu_1+\nu_2+\nu_3-\frac{D}{2}\right) \Gamma\left(\frac{D}{2}-\nu_2-\nu_3\right)}{\Gamma\left(\nu_1\right) \Gamma\left(\frac{D}{2}\right)} \nonumber\\
& \times S_1\left(\nu_2, \nu_1+\nu_2+\nu_3-\frac{D}{2}, \nu_3, 1+\nu_2+\nu_3-\frac{D}{2}, \frac{D}{2},-\frac{Q_1^2}{M_1^2}, \frac{M_2^2}{M_1^2}\right) \text {, } \\
& I_3^{\left\{p_1, q_1\right\}}=(-1)^{\frac{D}{2}}\left(-M_1^2\right)^{-\nu_1}\left(-M_2^2\right)^{\frac{D}{2}-\nu_2-\nu_3} \frac{\Gamma\left(\nu_2+\nu_3-\frac{D}{2}\right) \Gamma\left(\frac{D}{2}-\nu_3\right)}{\Gamma\left(\nu_2\right) \Gamma\left(\frac{D}{2}\right)} \nonumber\\
& \times H_2\left(\nu_2+\nu_3-\frac{D}{2}, \nu_3, \nu_1, \frac{D}{2}-\nu_3, \frac{D}{2}, \frac{Q_1^2}{M_2^2},-\frac{M_2^2}{M_1^2}\right)
\end{align}
Here, the $\nu_i$'s denote the powers of the propagators and $S_1$ is KdF function and  Horn $H_2$ function. In terms of our notation of KdF functions (see Appendix \ref{sec:definitionMHFs})
\begin{align*}
    S_1(a_1,a_2,b,c,d,x,y) = \text{KdF}^{2:1;0}_{1:1;0}
  \left[
   \setlength{\arraycolsep}{0pt}% local assignment
   \begin{array}{c@{{}:{}}c@{;{}}c}a_1, a_2& b & \linefill
 \\[1ex]
  c& d &  \linefill
   \end{array}
   \;\middle|\;
 x,y
 \right] 
\end{align*}

To proceed, we set the dimension $D = 4-2\e$ and the powers of the  propagators to unity (i.e. $\nu_i = 1, i = 1,2,3$). The KdF function $S_1$ takes a simpler form 
\begin{align} \label{eqn:application2F1}
    S_1(1,\e+1,1,\e+1,2-\e,x,y) &= \sum_{m,n=0}^\infty \frac{(1)_m  (1)_{m+n}}{ (2-\e)_m} \frac{x^m y^n}{m! n!}\nonumber\\
    &= -\frac{1}{y-1} \, _2F_1\left(1,1;2-\e;-\frac{x}{y-1}\right)
\end{align}
The second identity is obtained by performing the summation of the index $n$.

On the other hand, the Horn $H_2$ function takes the form 
\begin{align} \label{eqn:applicationH2}
    H_2(\e,1,1,1-\e,2-\e,x,y)
\end{align}

We can easily find the series expansion of the Gauss $_2F_1$ (\eeqref{eqn:application2F1}) and Horn $H_2$ function (\eeqref{eqn:applicationH2}) in $\e$ using the presented package (see Section \ref{sec:usage}), which reads
\begin{align} \label{eqn:expansionS1H2}
     S_1(1,\e+1&,1,\e+1,2-\e,x,y)
     = -\frac{1}{x} G\left(1;\frac{x}{1-y}\right)\nonumber\\
     &+\frac{\e}{x}\Bigg[G\left(1;\frac{x}{1-y}\right)-G\left(0,1;\frac{x}{1-y}\right)+G\left(1,1;\frac{x}{1-y}\right)\Bigg]+O\left(\e^2\right)\\
     H_2(\e,1,&1,1-\e,2-\e,x,y) = -\frac{G\left(\frac{1}{y}+1;x\right)}{x y}\nonumber\\
     &+\frac{\e }{x y}\Bigg(G\left(\frac{y+1}{y};x\right)-G\left(0,\frac{y+1}{y};x\right)\nonumber\\
     &+G\left(\frac{y+1}{y},1;x\right)+G\left(\frac{y+1}{y},\frac{y+1}{y};x\right)\Bigg)+O(\e^2)\label{eqn:expansionH2}
\end{align}

Using the above two expressions, we find the series expansion of the Feynman integral in terms of MPLs as

\begin{small}
    \begin{align}
   I_3^D&(1,1,1 ; Q_1^2, 0,0, M_1^2, M_2^2, 0) = \frac{1}{Q_1^2} \Bigg[G\left(0,1;-\frac{Q_1^2}{M_1^2-M_2^2}\right)-G\left(0,1-\frac{M_1^2}{M_2^2};\frac{Q_1^2}{M_2^2}\right)\\
   &-G\left(1,1;-\frac{Q_1^2}{M_1^2-M_2^2}\right)+G\left(1-\frac{M_1^2}{M_2^2},1;\frac{Q_1^2}{M_2^2}\right)+G\left(1-\frac{M_1^2}{M_2^2},1-\frac{M_1^2}{M_2^2};\frac{Q_1^2}{M_2^2}\right)\\
   &-\left(\log \left(-M_1^2\right)+\gamma +i \pi \right) G\left(1;-\frac{Q_1^2}{M_1^2-M_2^2}\right)+\left(\log \left(-M_2^2\right)+\gamma +i \pi \right) G\left(1-\frac{M_1^2}{M_2^2};\frac{Q_1^2}{M_2^2}\right)\Bigg]
\end{align}
\end{small}

which can further be written in terms of ordinary polylogarithms
as 
\begin{align}\label{eqn:example_series_expansion}
     I_3^D(1,1,1 ; Q_1^2, 0,0, M_1^2, &M_2^2, 0) = \frac{1}{Q_1^2} \Bigg( \text{Li}_{1,1}\left( 1- \frac{M_1^2}{M_2^2}, - \frac{Q_1^2}{M_1^2 - M_2^2}\right)\nonumber\\
     &+ \text{Li}_1\left(- \frac{Q_1^2}{M_1^2 - M_2^2}\right) \Big[\text{Li}_1\left(1+ M_2^2\right)-\text{Li}_1\left(1+ M_1^2\right)\Big]\Bigg)+ O(\e)
\end{align}

 The result obtained above in \eeqref{eqn:example_series_expansion} matches with Eq. (3.34) of \cite{Anastasiou:1999ui}  by analytic continuations. Thus we conclude that the series expansion coefficients of different analytic continuations of a Feynman integral are related by analytic continuations.   In other words, the process of finding analytic continuation and the process of obtaining the series expansion  of MHF commute with each other.

\section{Documentation and usage}
\label{sec:demo}

In this Section, we discuss the documentation and usage of the package \texttt{MultiHypExp}, which can be downloaded from the following url :

\begin{center}
    \href{https://github.com/souvik5151/MultiHypExp}{https://github.com/souvik5151/MultiHypExp}
\end{center}

It is built in \textsc{Mathematica} v11.3 and works in higher versions of \textsc{Mathematica} as well. The package depends on the following packages, whose usages are explained in Section \ref{sec:mathematica}.
\begin{itemize}
    \item  \href{http://www3.risc.jku.at/research/combinat/software/ergosum/RISC/HolonomicFunctions.html}{\texttt{HolonomicFunctions}} \cite{KoutschanPhD,koutschan2010fast}
    \item \href{https://github.com/christophmeyer/CANONICA}{\texttt{CANONICA}} \cite{Meyer:2017joq}
    \item \href{https://sites.google.com/site/loopcalculations/}{\texttt{HYPERDIRE}} \cite{Hyperdire,Hyperdire2,Hyperdire3}
    \item \href{https://gitlab.com/pltteam/plt/-/tree/master}{\texttt{PolyLogTools}} \cite{Duhr:2019tlz}
    \item \href{https://www.physik.uzh.ch/data/HPL/}{\texttt{HPL}} \cite{Maitre2006,Maitre:2007kp}
\end{itemize}

Therefore, those packages must be loaded before loading \texttt{MultiHypExp}. To do so, we store the path of the dependencies in the following global variables,

\begin{tcolorbox}
\begin{verbatim}
In[]:= $HYPERDIREPath = "Path_for_HYPERIDRE";
       $CANONICAPath =  "Path_for_CANONICA";
       $RISCPath = "Path_for_RISC_HolonomicFunctions";
       $PolyLogToolsPath = "Path_for_PolyLogTools";
       $HPLPath = "Path_for_HPL";    
\end{verbatim}    
\end{tcolorbox}

The \texttt{CANONICA} and \texttt{HYPERDIRE} packages are loaded inside the package \texttt{MultiHypExp} and the \texttt{HPL} package is called automatically when \texttt{PolyLogTools} is called.  The user must load the other dependencies after setting their paths.

\begin{tcolorbox}
\begin{verbatim}
In[]:= $Path = Join[$Path, {$PolyLogToolsPath, $HPLPath, $RISCPath}];
       << PolyLogTools`;
       << RISC`HolonomicFunctions`;
\end{verbatim}    
\end{tcolorbox}

The performance of the package \texttt{MultiHypExp} can be significantly improved by calling the \texttt{PolyLogTools} package in multiple kernels. This can be achieved by distributing the paths of \texttt{PolyLogTools} and \texttt{HPL} in the available kernels.

\begin{tcolorbox}
\begin{verbatim}
In[]:= LaunchKernels[2];
       ParallelEvaluate[$Path = Join[$Path, {$PolyLogToolsPath,
       $HPLPath}]];
       ParallelNeeds["PolyLogTools`"];
\end{verbatim}    
\end{tcolorbox}

Finally, the package \texttt{MultiHypExp} can be loaded after setting its path.

\begin{tcolorbox}
\begin{verbatim}
In[]:= AppendTo[$Path, "Path_for_MultiHypExp"];
       << MultiHypExp.wl  
\end{verbatim}    
\end{tcolorbox}

We now demonstrate the usage of the commands.
The package \texttt{MultiHypExp} consists of two commands, \texttt{SeriesExpand} and \texttt{ReduceFunction}.

\subsection{The \texttt{SeriesExpand}  command}
\label{sec:commandseriesexpand}
The command \texttt{SeriesExpand} finds the series expansion of MHFs. It can take input in two different forms, which we discuss below.

\begin{tcolorbox}
\texttt{SeriesExpand[FunctionName,Parameter\_List,Var\_List,e\_Symbol,p\_Integer]}
\end{tcolorbox}
or,
\begin{tcolorbox}
\textbf{\texttt{SeriesExpand[Indices\_List,exp,Var\_List,e\_Symbol,p\_Integer]}}
\end{tcolorbox}

This command finds the first \texttt{p\_Integer} coefficients of the series expansion of the function \texttt{FunctionName} in \texttt{e\_Symbol}. The other input arguments of the commands are given below.
\begin{itemize}
    \item \texttt{FunctionName} : The name of the HF. The following are the available \texttt{FunctionName}'s of double- and triple-variable HFs.\\
    Double variable series : F1, F2 ,F3, F4, G1, G2, G3, H1, H2, H3, H4,
H6 and H7\\
    Three variable series : FA3, FB3, FD3, FK3, FM3, FN3 and FS3
    \item \texttt{Parameter\_List} : List of Pochhammer parameters
    \item \texttt{Var\_List} : List of variables
    \item \texttt{exp} : Expression of the MHF
    \item \texttt{Indices\_List}  : List of summation indices
    \item \texttt{p\_Integer} : asked number of series coefficients
    \item \texttt{e\_Symbol} : the series expansion parameter, which must be a symbol  
\end{itemize}

Note that, for the expansion of $_pF_{p-1}$ functions, the \texttt{FunctionName} input must be omitted and the \texttt{Parameter\_List} argument must be given in the following form\\
\texttt{ Parameter\_List\\
= \{Upper\_Pochhammer\_Parameter\_List, Lower\_Pochhammer\_Parameter\_List\}}

\subsection{The \texttt{ReduceFunction} command}
\label{sec:ReduceFunctioncommand}

This command finds the reduction formulae of MHFs in terms of MPLs. It takes the input as

\begin{tcolorbox}
    \texttt{ReduceFunction[FunctionName, Parameter\_List, Var\_List]}
\end{tcolorbox}
where, as before
\begin{itemize}
    \item  \texttt{Parameter\_List} : List of Pochhammer parameters
    \item \texttt{Var\_List} : List of variables
    \item \texttt{FunctionName} : The name of the HF. The available \texttt{FunctionName}'s are\\
    Double variable series : F1, F2, F3 and F4\\
    Triple variable series : FD3 and FS3
\end{itemize}

We next show the usage of the commands by reproducing the results of the series expansion of the MHFs considered in Section \ref{sec:examples} and Section \ref{sec:applications}.

\subsection{Usage of the commands}
\label{sec:usage}

In this Section, we provide demonstrations of the two commands of our package.

\subsubsection{ The \texttt{SeriesExpand} command}
Let us consider the one variable Gauss $_2F_1$ function \eeqref{eqn:examples2f1def} from Section \ref{sec:examples2f1},

\begin{align*}
     F(\e) := ~_2F_1(\e,-\e;\e-1;x) =  \sum_{m=0}^\infty \frac{ (\e)_m (-\e)_m}{ (\e-1)_m} \frac{x^m}{m!}  
\end{align*}

Below we find the series expansion of the above function using the  \texttt{SeriesExpand} command.

\begin{tcolorbox}
\begin{verbatim}
In[]:= SeriesExpand[{{e, -e}, {e - 1}}, {x}, e, 3]

Out[]= 1 + (-(x/(-1+x))+G[1,x]) e
    + (-(x/(-1+x))-(x G[1,x])/(-1+x)+G[1,1,x]) e^2 + O[e]^3
\end{verbatim}
\end{tcolorbox}

The same result can be obtained by providing the series representation of the Gauss $_2F_1$ function in the \texttt{SeriesExpand} command as follows.

\begin{tcolorbox}
\begin{verbatim}
In[]:= SeriesExpand[{n},(Pochhammer[e,n]Pochhammer[-e,n]x^n)
/(Pochhammer[e-1,n]n!),{x},e,3] 
\end{verbatim}
\end{tcolorbox}

We now produce the series expansion of bi-variate Appell $F_2$ functions considered in Section \ref{sec:examplesF2}. Using the \texttt{SeriesExpand} command, we find the series expansion of $F'$ (defined in \eqref{eqn:examplesF2def2}) to be

\begin{tcolorbox}
\begin{verbatim}
In[]:= SeriesExpand[F2,{1,1,e,1+e,1-e},{x,y},e,3]

Out[] = -(1/(-1+x))+((-2 G[1,x]+G[1,y]+G[1-y,x]) e)/(-1+x)
+(1/(-1+x))(2 G[1,x] G[1,y]-2 G[1,y] G[1-y,x]+2 G[0,1,x]+G[0,1,y]
-G[0,1-y,x]-4 G[1,1,x]-2 G[1,1,y]+2 G[1,1-y,x]+2 G[1-y,1,x]
-G[1-y,1-y,x]) e^2+O[e]^3
\end{verbatim}
\end{tcolorbox}
which matches with \eeqref{eqn:Fpexpansion}. 

In a similar fashion, we call the \texttt{SeriesExpand} command to yield the series expansion of $F$ (defined in \eeqref{eqn:examplesF2def1}),

\begin{tcolorbox}
\begin{verbatim}
In[]:= SeriesExpand[F2,{1,1,e,+e,-e},{x,y},e,2]

Out[]= (x (2/(-1+x)^2-1/(-1+x+y)^2))/e
+(-(((-1+x+2 y) (1-y+x (-3+2 x+3 y)))/((-1+x)^2 (-1+x+y)^2))
+x (4/(-1+x)^2-2/(-1+x+y)^2) G[1,x]
+2 x (-(1/(-1+x)^2)+1/(-1+x+y)^2) G[1,y]
+x (-(2/(-1+x)^2)+1/(-1+x+y)^2) G[1-y,x])+O[e]^1
\end{verbatim}
\end{tcolorbox}
Thus, we reproduce \eqref{eqn:Fexpansion}.
Likewise, the series expansion of the Horn $H_2$ function in Section \ref{sec:applications} (\eeqref{eqn:expansionH2}) can be obtained as

\begin{tcolorbox}
\begin{verbatim}
In[]:= SeriesExpand[H2,{e,1,1,1-e,2-e},{x,y},e,2]

Out[]= -(G[1+1/y,x]/(x y))+((G[(1+y)/y,x]-G[0,(1+y)/y,x]
+G[(1+y)/y,1,x]+G[(1+y)/y,(1+y)/y,x]) e)/(x y)+O[e]^2
\end{verbatim}
\end{tcolorbox}

Finally, we provide an example of a series expansion of a HF using the transformation theory. As an example, we take the Horn function $G_2$ with the following set of Pochhammer parameters.

\begin{align}
    G_2:= G_2 (\e,\e,\e,\e;x,y)
\end{align}

The series expansion of $G_2$ is calculated using its transformation formula to Appell $F_2$ (i.e., \eeqref{eqn:G2F2relation}) internally inside the package. Therefore, we call the command to find the series expansion of $G_2$ as 

\begin{tcolorbox}
\begin{verbatim}
In[]:= SeriesExpand[G2,{e,e,e,e},{x,y},e,2]

Out[]= 1+(-G[0,1+x]-G[0,1+y]-G[1,x/(1+x)]-G[1,y/(1+y)]) e+O[e]^2
\end{verbatim}
\end{tcolorbox}

\subsubsection{The \texttt{ReduceFunction} command}

As discussed in Section \ref{sec:ReduceFunctioncommand}, this command finds the reduction formula of the MHFs. We demonstrate the usage of this command with examples of double- and triple-variable HFs. 

\begin{tcolorbox}
\begin{verbatim}
In[]:= ReduceFunction[F2,{3,2,1,3,2},{x,y}]

Out[] = 1/((-1+x) x (-1+x+y))-G[1,x]/(x^2 y)+G[1-y,x]/(x^2 y)
\end{verbatim}
\end{tcolorbox}
Thus, we reproduce \eeqref{eqn:F2reduction}. 

We conclude this section by presenting an example of the reduction formula of  the triple variable Lauricella-Saran function $F_S(1,1,1,1,1;2;x,y,z)$. 
\begin{tcolorbox}
\begin{verbatim}
In[]:= ReduceFunction[FS3,{1,1,1,1,1,2},{x,y,z}]

Out[]= -((x G[1,x])/((x (-1+y)-y) (x (-1+z)-z)))
+(y G[1,y])/((x (-1+y)-y) (y-z))-(z G[1,z])/((y-z) (x (-1+z)-z))
\end{verbatim}
\end{tcolorbox}

which matches \eeqref{eqn:FSreduction}.

\section{\textsc{Mathematica} implementation}
\label{sec:mathematica}

The presented method in Section \ref{sec:method} is implemented in the form of a \textsc{Mathematica} \cite{Mathematicav11.3} package \texttt{MultiHypExp}, which consists of two commands, \texttt{SeriesExpand} and \texttt{ReduceFuntion}. The former command can be used to find the series expansion of a given MHF, and the latter can be used to yield reduction formulae of MHFs. The usage of these commands is demonstrated in Section \ref{sec:demo} in detail. In this Section, we discuss how these commands are implemented in \textsc{Mathematica}. 

\subsection{The \texttt{SeriesExpand} command}

Let us start with the \texttt{SeriesExpand} command. In Section \ref{sec:method}, the procedure of finding the series expansion is divided into five parts.

\begin{itemize}
    \item \textit{Step 1} and \textit{Step 3} : In the first step (Section \ref{sec:step1}), the type of the series expansion of the given MHF is determined by its Pochhammer structure, and the singular Pochhammer parameters (if exist) are replaced by the non-singular ones according to the prescription given in \textit{Step 3} (Section \ref{sec:step3}). This is achieved using the \textsc{Mathematica}'s pattern matching and replacement rules commands.
    
    \item \textit{Step 2} : The Taylor series of MHF is found in this step. At first, the PDE associated with the given MHF is obtained and brought to the Pfaffian form. 
    
    \begin{enumerate}
        \item When the \texttt{SeriesExpand} command is called with the input argument\\ \texttt{FunctionName} (i.e., the first form in Section \ref{sec:commandseriesexpand}), the pre-stored Pfaff systems of double variable Appell $F_1, F_2, F_3$ and Horn $H_2$ and triple variable Lauricella $F_A, F_B, F_D, F_K$, $F_M$, $F_N$, and $F_S$ are used. 
        \item  When the \texttt{SeriesExpand} command is called with the input argument \texttt{exp} (i.e., in the second form in Section \ref{sec:commandseriesexpand}), the PDE associated with the MHF is calculated using the expressions given in \eeqref{eqn:annihilator1} and \eeqref{eqn:annihilator2}. Further, the system of PDE is brought to Pfaffian form using the Gr\"obner basis calculations, which is carried out by the \textsc{Mathematica} package \texttt{HolonomicFunctions} \cite{KoutschanPhD,koutschan2010fast}. It may be noted that, there are computer programs which allow one to perform the Gr\"obner basis calculations, e.g., Singular \cite{DGPS}, Macaulay2 \cite{M2}, Maple \cite{Chyzak1998}. However, we opt to utilize \texttt{HolonomicFunctions} in order to maintain the  homogeneity of the package within the scope of \textsc{Mathematica}.
    \end{enumerate}

    Then the Pfaffian system, wherever possible, is further brought to the canonical form using the \textsc{Mathematica} package \texttt{CANONICA} \cite{Meyer:2017joq} and solved using proper boundary condition and expressed in terms of MPLs. Extensive use of commands of \texttt{PolyLogTools} \cite{Duhr:2019tlz} are made to integrate and simplify the expressions containing MPLs.

    \item \textit{Step 4} : The differential operator that relates the given MHF and the associated secondary function is found in this step. The \texttt{HYPERDIRE} \cite{Hyperdire,Hyperdire3} packages are used to find the necessary step up/down operators for the Appell $F_1, F_2, F_3$ and Horn $H_2$ functions when the \texttt{SeriesExpand} command is called using the input argument \texttt{FunctionName}. Otherwise, the operators are calculated from \eeqref{eqn:stepdownop} and \eeqref{eqn:stepupop} and the reduction of the products of these step up/down operators with respect to the Gr\"obner basis of the annihilators is performed using the \texttt{HolonomicFunctions} package.

    \item \textit{Step 5} : Finally, the obtained differential operator is made to act on the Taylor series of the secondary function. As the Taylor expansion coefficients contain MPLs, the \texttt{PolyLogTools} commands are used to calculate the derivatives and simplify those expressions.

\end{itemize}

The series expansion of other double variable Appell-Horn functions (i.e., $F_4, H_1, H_3$,$ H_4$, $ H_6$ and $H_7$) are calculated using their connection formulae to Appell $F_1, F_2, F_3$ or Horn $H_2$, which are given in Appendix \ref{sec:Transformation_Theory}.

\subsection{The \texttt{ReduceFunction} command}

It is possible to find the reduction formulae of MHFs as a byproduct of the methodology. For a given  MHF, one can find  a secondary MHF, whose Taylor expansion can be readily calculated and the differential operator associated with these two functions can be made to act on the series expansion of the secondary function to find the  reduction of the given MHF. An example of such reduction of Appell $F_2$ function is given in Appendix \ref{sec:Reduction_formulae}. This procedure is implemented in the  \texttt{ReduceFunction} command, which can find the reduction formulae of double variable Appell $F_1$, $F_2$, $F_3$, $F_4$ and triple variable Lauricella $F_D$ and $F_S$ functions. The required differential operators are procured using the \texttt{HYPERDIRE} \cite{Hyperdire, Hyperdire2} packages, and the Taylor expansions of the abovementioned MHFs from Appendix \ref{sec:exactresult} are stored in the package and  are called accordingly to fulfill the purpose. It should be noted that, the current version of the package does not provide reduction formulae for other Horn or Lauricella functions.

\subsection{Comments on the implementation }
This section addresses various observations concerning the current version of the \textsc{Mathematica} implementation.
\begin{enumerate}
    \item In principle, the procedure mentioned in \cite{Bera:2022ecn} can be applied to find the series expansion of any MHFs. However, the package \texttt{MultiHypExp} is made with the objective of expressing the series coefficients in terms of well-known MPLs. This, not only restricts the reach of the package to the small number of double- and triple-variable HFs, but also restricts the Pochhammer parameters to integers for most of the accessible HFs. Thus, the  Appell $F_1, F_2, F_3$, and Horn $H_2$ and triple variable Lauricella $F_A$, $F_B$, $F_D$, $F_K$, $F_M$, $F_N$, and $F_S$ functions can be expanded in series around the integer values of their Pochhammer parameters.
    \item Since the series expansion of other double variable Appell-Horn functions (i.e., $F_4$, $ H_1$, $ H_3$, $ H_4$, $ H_6$ and $H_7$) are calculated using their connection formulae to Appell $F_1, F_2, F_3$ or Horn $H_2$, the parameter $c$ in $H_4$, $d$ in $H_7$ must be of the form $ p/2 + q \e$, where $p$ is integer or half integer, so that, the $F_2$ and $H_2$ function appearing on the right-hand side of their corresponding connection formula (see Appendix \ref{sec:Transformation_Theory}), can be expanded in series around integer values of parameters.
    \item To the best of our knowledge, there do not exist connection formulae of Appell $F_4$, Horn $H_1$ and $H_5$ with general parameters to other Appell-Horn functions. Thus, the package can find the series expansion of Appell $F_4$ and Horn $H_1$ with restricted Pochhammer parameters (see Appendix \ref{sec:Transformation_Theory}).
    \item The package in its current version can find, at most, the first six series coefficients. This limitation comes from the \texttt{ToFibrationBasis} command of \texttt{PolyLogTools} as it can handle MPLs with weight up to five. This restriction can be lifted with the introduction of the newer version of \texttt{PolyLogTooLs} when it becomes available.
%    \item In certain situations, a non-rational transformation of variables is required to bring the Pfaffian system to the canonical form. This is neither implemented in \texttt{CANONICA} nor in \texttt{MultiHypExp}. However, 
    \item In certain situations, a non-linear change of variables is required to rationalize algebraic functions appearing in  the Pfaffian system, which has not been implemented in the present version of the package. This topic has been studied in the literature (see e.g., \cite{Besier:2018jen}).  We intend to incorporate these into our future work.

    \item On an ordinary personal computer, the process of finding series expansion takes a few minutes for  double variable HFs. However, the computational time significantly increases when triple variable HFs are considered, as most of the time is devoted to bringing the Pfaffian system into canonical form.
    \item The series expansion of a given MHF is not applicable on its singular locus.
    \item Since the series expansion of the Appell $F_4$ function with general values of Pochhammer parameters can not be found using this package, the \texttt{ReduceFunction} command can only find the reduction formulae of $F_4$ for restricted parameters.
\end{enumerate}

\section{Summary and Conclusions }
\label{sec:Conclusions}

In this work, we have provided an implementation of an algorithm to perform series expansion of MHFs. The implemented algorithm is a slight modification of the one presented in \cite{Bera:2022ecn}. The modifications are made for the requirement of expressing the series coefficients in terms of MPLs, rather than higher summation-fold MHFs. We have utilized publicly available packages as dependencies to build the package \texttt{MultiHypExp}, which, in its  current version, is suitable for finding the series expansion of certain one, two, and three-variable HFs around the integer values of parameters. The restriction of the (integer-valued) parameters arises as we choose to express the series coefficients in terms of MPLs, which offers immediate numerical evaluation using well-established computer programs. We have described the steps of the algorithms in detail and explained how the dependencies are employed to accomplish those steps.  
We have provided examples of the series expansion of MHFs that typically appears in Feynman integral calculus. 
The package also allows one to find reduction formulae of certain MHFs with integer values of Pochhammer parameters to simpler functions.

%The package can be easily integrated with other publicly available packages like \texttt{FeynGKZ} \cite{Ananthanarayan:2022ntm}, \texttt{Olsson} \cite{Olssonwl}, which find the HF representation of a Feynman integral using GKZ techniques and the analytic continuations of MHFs respectively. This will allow to to 

In some cases, the MHFs are needed to series-expand around rational values of parameters, which may require functions beyond MPLs. The current version of the package is not suitable for that. We plan to explore the possibilities in near future.

\acknowledgments
We are indebted to Prof. Thomas Gehrmann and the Physik-Institut, Universit\"at Z\"urich for supporting the present work and hospitality.  We are also grateful to Prof. Daniel Wyler for his support throughout the course of this work. We thank Prof. Thomas Gehrmann, Dr. Robin Marzucca and Dr. Kay Sch\"onwald for enlightening discussions and Prof. B. Ananthanarayan and  Prof. Thomas Gehrmann for useful comments on the manuscript. This is a part of the author's doctoral work at CHEP, IISc.

\appendix
\section{Definition of Multiple polylogarithms (MPLs)}

\label{sec:definitionMPL}

Multiple polylogarithms \cite{goncharov2001multiple, goncharov2011multiple}  are defined as
\begin{align}
    G(a_1,\dots , a_n;z) = \int_0^z \frac{dt}{t-a_1} G(a_2,\dots,a_n;t )
\end{align}
with $G(z) = 1$ and $a_i$ and $z$ are complex-valued variables. The vector $\vec{a} = (a_1, \dots ,a_n)$ is called the weight vector, and its length is called the weight.

These can also be defined as nested sums,

\begin{align}
\text{Li}_{m_1, \ldots, m_k}\left(z_1, \ldots, z_k\right) & =\sum_{0<n_1<n_2<\cdots<n_k} \frac{z_1^{n_1} z_2^{n_2} \cdots z_k^{n_k}}{n_1^{m_1} n_2^{m_2} \cdots n_k^{m_k}} \\
& =\sum_{n_k=1}^{\infty} \frac{z_k^{n_k}}{n_k^{m_k}} \sum_{n_{k-1}=1}^{n_k-1} \ldots \sum_{n_1=1}^{n_2-1} \frac{z_1^{n_1}}{n_1^{m_1}},
\end{align}

which are related to each other as

\begin{align}
    G\left(\vec{0}_{m_1-1}, a_1, \ldots, \vec{0}_{m_k-1}, a_k ; z\right) &=(-1)^k~ \text{Li}_{m_k, \ldots, m_1}\left(\frac{a_{k-1}}{a_k}, \ldots, \frac{a_1}{a_2}, \frac{z}{a_1}\right)\\
G\left(\vec{a}_n ; z\right) & =\frac{1}{n !} \log ^n\left(1-\frac{z}{a}\right) \label{eqn:MPL2Log}, \\
G\left(\vec{0}_{n-1}, a ; z\right) & =-\text{Li}_n\left(\frac{z}{a}\right),
\end{align}

where $a_i \neq 0$. In \eeqref{eqn:MPL2Log}, we assume $\vec{a}_n$ contains $n$ equal, nonzero entries $a$.

MPLs, GPLs, logarithms and polylogarithms are used interchangeably throughout the paper.

\section{Definitions of some MHFs}
\label{sec:definitionMHFs}

In the appendix, we list the definition of some MHFs in two and three variables along with their domain of convergences below. The standard references are \cite{Appell1880, Bateman:1953, Slater:1966,Exton:1976, Bailey:1964, Srivastava:1985}.

The four Appell functions are defined as
\begin{align}
    &F_1 := F_{1}\left(a, b_{1}, b_{2} ; c ; x, y\right)=\sum_{m, n=0}^{\infty} \frac{(a)_{m+n}\left(b_{1}\right)_{m}\left(b_{2}\right)_{n}}{(c)_{m+n}}  \frac{ x^{m} y^{n}}{ m ! n !}\\
    &F_2 := F_{2}\left(a, b_{1}, b_{2} ; c_{1}, c_{2} ; x, y\right)=\sum_{m, n=0}^{\infty} \frac{(a)_{m+n}\left(b_{1}\right)_{m}\left(b_{2}\right)_{n}}{\left(c_{1}\right)_{m}\left(c_{2}\right)_{n}} \frac{ x^{m} y^{n}}{ m ! n !}\\
    &F_3 :=F_{3}\left(a_{1}, a_{2}, b_{1}, b_{2} ; c ; x, y\right)=\sum_{m, n=0}^{\infty} \frac{\left(a_{1}\right)_{m}\left(a_{2}\right)_{n}\left(b_{1}\right)_{m}\left(b_{2}\right)_{n}}{(c)_{m+n}} \frac{ x^{m} y^{n}}{ m ! n !}\\
    &F_4 :=F_{4}\left(a, b ; c_{1}, c_{2} ; x, y\right)=\sum_{m, n=0}^{\infty} \frac{(a)_{m+n}(b)_{m+n}}{\left(c_{1}\right)_{m}\left(c_{2}\right)_{n} }  \frac{ x^{m} y^{n}}{ m ! n !}
\end{align}
their associated domains of convergence are
\begin{align}
    F_1 &: |x|<1 \wedge |y|<1\\
    F_2 &: |x|+|y|<1\\
    F_3 &: |x|<1 \wedge |y|<1\\
    F_4 &: \sqrt{|x|}+\sqrt{|y|}<1
\end{align}

The ten double-variable Horn functions are defined below.

\begin{align}
    &G_1 := G_1\left(a , b, b^{\prime} ; x, y\right)=\sum_{m, n=0}^{\infty}(a)_{m+n}(b)_{n-m}\left(b^{\prime}\right)_{m-n} \frac{x^m y^n}{m ! n !} \\
    &G_2 := G_2\left(a, a^{\prime} , b, b^{\prime} ; x, y\right)=\sum_{m, n=0}^{\infty}(a)_m\left(a^{\prime}\right)_n(b)_{n-m}\left(b^{\prime}\right)_{m-n} \frac{x^m y^n}{m ! n !} \\
    &G_3 := G_3\left(a, a^{\prime} ; x, y\right)=\sum_{m, n=0}^{\infty}(a)_{2 n-m}\left(a^{\prime}\right)_{2 m-n} \frac{x^m y^n}{m ! n !} \\
    &H_1 := H_1(a , b , c ; d ; x, y)=\sum_{m, n=0}^{\infty} \frac{(a)_{m-n}(b)_{m+n}(c)_n}{(d)_m} \frac{x^m y^n}{m ! n !} \\
    &H_2 := H_2(a , b , c , d ; e ; x, y)=\sum_{m, n=0}^{\infty} \frac{(a)_{m-n}(b)_m(c)_n(d)_n}{(e)_m} \frac{x^m y^n}{m ! n !} \\
    &H_3 := H_3(a , b ; c ; x, y)=\sum_{m, n=0}^{\infty} \frac{(a)_{2 m+n}(b)_n}{(c)_{m+n}} \frac{x^m y^n}{m ! n !} \\
    &H_4:= H_4(a , b ; c , d ; x, y)=\sum_{m, n=0}^{\infty} \frac{(a)_{2 m+n}(b)_n}{(c)_m(d)_n} \frac{x^m y^n}{m ! n !} \\
    &H_5:= H_5(a , b ; c ; x, y)=\sum_{m, n=0}^{\infty} \frac{(a)_{2 m+n}(b)_{n-m}}{(c)_n} \frac{x^m y^n}{m ! n !} \\
    &H_6:= H_6(a , b , c ; x, y)=\sum_{m, n=0}^{\infty}(a)_{2 m-n}(b)_{n-m}(c)_n \frac{x^m y^n}{m ! n !} \\
    &H_7:= H_7(a , b , c ; d ; x, y)=\sum_{m, n=0}^{\infty} \frac{(a)_{2 m-n}(b)_n(c)_n}{(d)_m} \frac{x^m y^n}{m ! n !} 
\end{align}
The associated domains of convergence are
\begin{align}
    G_1 &: |x|+|y|<1\\
    G_2 &: |x|<1 \wedge |y|<1\\
    G_3 &: Z_1 \cap Z_2, \hspace{1cm}Z_1 = {|x|<\Phi_1(|y|)}, Z_2 = {|y|<\Phi_1(|x|)}\\
    H_1 &: \left| x\right| <1\land \left| y\right| <1\land 2 \sqrt{\left| x\right|  \left| y\right| }+\left| y\right| <1\\
    H_2 &: \left| x\right| <1\land \left| y\right| <1\land|y|(1+|x|)<1\\
    H_3 &: \left| x\right| <\frac{1}{4}\wedge \left( |x|<\frac{1}{4}\wedge |y|< \frac{1}{2} + \frac{1}{2}\sqrt{1-4 |x|} \right) \cup \left(|y|\leq \frac{1}{2}\right)\\
    H_4 &: 2\sqrt{|x|}+ |y|<1\\H_5 &: |x|<\frac{1}{4}\wedge |y|< \text{min}\{\Psi_1(|x|),\Psi_2(|x|)\}
\end{align}

\begin{align}   
    H_6 &: \left| x\right| <\frac{1}{4}\wedge \left(|x| |y|^2 + |y|<1\right)\\
    H_7 &: \left| x\right| <\frac{1}{4}\wedge |y|(1+ 2\sqrt{|x|})<1 
\end{align}

where
\begin{align}
    \Phi_1(x) &= \frac{2 \sqrt{3 x+1}+1}{3 \left(\sqrt{3 x+1}+1\right)^2}\\
    \Phi_2(x) &= \frac{2 \sqrt{1-3 x}-1}{3 \left(\sqrt{1-3 x}-1\right)^2}\\
    \Psi_1(x) &=\frac{2 \left(2-\sqrt{12 x+1}\right)^2}{9 \left(\sqrt{12 x+1}-1\right)}\\
    \Psi_2(x) &= \frac{2 \left(\sqrt{1-12 x}+2\right)^2}{9 \left(\sqrt{1-12 x}+1\right)}
\end{align}
These fourteen Appell-Horn functions form the set of complete, order two bivariate HFs.

The Kamp\'e de F\'eriet functions are defined as
\begin{small}
    \begin{align}
\text{KdF}^{p:q;k}_{l:m;n}
  \left[
   \setlength{\arraycolsep}{0pt}% local assignment
   \begin{array}{c@{{}:{}}c@{;{}}c} (a_p)&(b_q)&(c_k)
 \\[1ex]
  (\alpha_l)&(\beta_m)& (\gamma_n) 
   \end{array}
   \;\middle|\;
 x,y
 \right] := \sum_{r=0}^{\infty} \sum_{s=0}^{\infty} \frac{\prod_{j_1=1}^p\left(a_{j_1}\right)_{r+s} \prod_{j_2=1}^q\left(b_{j_2}\right)_r \prod_{j_3=1}^k\left(c_{j_3}\right)_s}{\prod_{j_4=1}^l\left(\alpha_{j_4}\right)_{r+s} \prod_{j_5=1}^m\left(\beta_{j_5}\right)_r \prod_{j_6=1}^n\left(\gamma_{j_6}\right)_s} \frac{x^r}{r !} \frac{y^s}{s !}
\end{align}

\end{small}

and

\begin{small}
    \begin{align}
\widetilde{\text{KdF}}^{p:q;k}_{l:m;n}
  \left[
   \setlength{\arraycolsep}{0pt}% local assignment
   \begin{array}{c@{{}:{}}c@{;{}}c} (a_p)&(b_q)&(c_k)
 \\[1ex]
  (\alpha_l)&(\beta_m)& (\gamma_n) 
   \end{array}
   \;\middle|\;
 x,y
 \right] := \sum_{r=0}^{\infty} \sum_{s=0}^{\infty} \frac{\prod_{j_1=1}^p\left(a_{j_1}\right)_{r-s} \prod_{j_2=1}^q\left(b_{j_2}\right)_r \prod_{j_3=1}^k\left(c_{j_3}\right)_s}{\prod_{j_4=1}^l\left(\alpha_{j_4}\right)_{r-s} \prod_{j_5=1}^m\left(\beta_{j_5}\right)_r \prod_{j_6=1}^n\left(\gamma_{j_6}\right)_s} \frac{x^r}{r !} \frac{y^s}{s !}
\end{align}
\end{small}

The domains of convergence of the KdF functions are,
\begin{enumerate}
    \item $p+q< l+m+1, p+k<l+n+1, |x|<\infty, |y|<\infty$\\
or,
    \item $p+q=l+m+1, p+k=l+n+1$ and\\
 \[
 \begin{cases}
        |x|^{1/(p-l)}+ |y|^{1/(p-l)} <1,& \hspace{.5cm} \text{if} \hspace{.5cm} p>l,\\
        \text{max}\{|x|,|y|\}<1,& \hspace{.5cm} \text{if}  \hspace{.5cm} p\leq l
    \end{cases}
\]
\end{enumerate}
The $\widetilde{\text{KdF}}$ functions can be related to the $\text{KdF}$ functions by reshuffling the summation indices (see C.2. of \cite{DelDuca:2009ac}). Thus, we do not discuss the domains of convergence of these functions. However, the domain of MHFs can be derived using the Horn's theorem, which is implemented in the \texttt{Olsson} \cite{Olssonwl} package for the bivariate HFs. 

Next, we list the triple-variable HFs whose series expansion can be performed by the \texttt{MultiHypExp} package.
\begin{small}

\begin{align}
    F_A :=&F_A(a,b_1,b_2,b_3;c_1,c_2,c_3; x,y,z) = \sum_{m,n,p = 0}^\infty\frac{ \left(a\right)_{m+n+p} (b_1)_m (b_2)_n (b_3)_p  }{(c_1)_m (c_2)_n (c_3)_p } \frac{x^m y^n z^p}{m! n! p!}\\
    F_B :=&F_B(a_1,a_2,a_3,b_1,b_2,b_3;c; x,y,z) = \sum_{m,n,p = 0}^\infty\frac{ (a_1)_m (a_2)_n (a_3)_p  (b_1)_m (b_2)_n (b_3)_p  }{(c)_{m+n+p}} \frac{x^m y^n z^p}{m! n! p!}\\
    F_D :=&F_D(a,b_1,b_2,b_3;c_1,c_2,c_3; x,y,z) = \sum_{m,n,p = 0}^\infty\frac{ \left(a\right)_{m+n+p} (b_1)_m (b_2)_n (b_3)_p  }{(c)_{m+n+p} } \frac{x^m y^n z^p}{m! n! p!}\\
    F_K :=&F_K(a_1,a_2,b_1,b_2;c_1,c_2,c_3; x,y,z) = \sum_{m,n,p = 0}^\infty\frac{ \left(a_1\right)_m \left(a_2\right)_{n+p}  \left(b_1\right)_{m+p} \left(b_2\right)_n }{ \left(c_1\right)_m \left(c_2\right)_n \left(c_3\right)_p} \frac{x^m y^n z^p}{m! n! p!}\\
    F_M :=&F_M(a_1,a_2,b_1,b_2;c_1,c_2;x,y,z) = \sum_{m,n,p = 0}^\infty \frac{ \left(a_1\right)_m  \left(a_2\right)_{n+p} \left(b_1\right)_{m+p}\left(b_2\right)_n}{ \left(c_1\right)_m \left(c_2\right)_{n+p}} \frac{x^m y^n z^p}{m! n! p!}\\
    F_N :=&F_N(a_1,a_2,a_3,b_1,b_2;c_1,c_2;x,y,z) = \sum_{m,n,p = 0}^\infty \frac{ \left(a_1\right)_m \left(a_2\right)_n \left(a_3\right)_p  \left(b_1\right)_{m+p}\left(b_2\right)_n}{ \left(c_1\right)_m \left(c_2\right)_{n+p}} \frac{x^m y^n z^p}{m! n! p!}\\
    F_S :=&F_S(a_1,a_2,b_1,b_2,b_3;c;x,y,z) = \sum_{m,n,p = 0}^\infty \frac{ \left(a_1\right)_m \left(a_2\right)_{n+p}   \left(b_1\right)_{m}\left(b_2\right)_n \left(b_2\right)_p}{ \left(c\right)_{m+n+p} } \frac{x^m y^n z^p}{m! n! p!}
\end{align}
    
\end{small}

The corresponding domains of convergence are
\begin{align}
    F_A &: |x|+|y|+|z|<1\\
    F_B &:  |x|<1 \wedge|y|<1\wedge |z|<1\\
    F_D &: |x|<1 \wedge|y|<1\wedge|z|<1\\
    F_K &: |x|<1 \wedge |z|<1 \wedge |y|<(1-|x|)(1-|z|)\\
    F_M &: |x|<1 \wedge |y|+|z|<1\\
    F_N &: |x|+ |y|<1 \wedge |z|<1 \\
    F_S &: |x|<1 \wedge|y|<1\wedge |z|<1\\
\end{align}

\section{Some results} \label{sec:exactresult}

We provide some series expansion of pure functions below. Note that, each of the series coefficients is of  weight zero if we assign the weight of $\e^n$ to be $-n$. The weight of $G(a_1,\dots,a_n;x)$ is $n$.

\begin{scriptsize}

\begin{align}
    F_1(a\e,&b_1 \e, b_2 \e; 1+ c \e; x, y) = 1+ \e^2 [ -a b_1 G(0,1,x)-a b_2 G(0,1,y) ]
    +\e^3 [a b_1 G(0,1,1,x) \left(a+b_1-c\right)\nonumber\\
    &+a b_1 \left(c-b_2\right) G(0,0,1,x)+a b_2 c G(0,0,1,y)+a b_2 G(0,1,1,y) \left(a+b_2-c\right)\nonumber\\
    &+a b_1 b_2 G(0,1,x) G(1,y)-a b_1 b_2 G(1,y) G(0,y,x)+a b_1 b_2 G(0,y,1,x)]  + O(\e^4)\\
    %%%%%%%%%%%%%%%%%%%
    %%%%%%%%%%%%%%%% F2 %%%%%%%
    F_2(a\e,&b_1 \e,b_2 \e;1+c_1 \e,1+c_2\e;x,y) = 1+ \e^2\left[ -a b_1 G(0,1,x)-a b_2 G(0,1,y) \right]\nonumber\\
    &+\e^3 [G(0,1,1,y) \left(a^2 b_2-a b_2 c_2+a b_2^2\right)+a b_1 (G(0,1,1,x) \left(a+b_1-b_2-c_1\right)\nonumber\\
    &+b_2 G(0,1,x) G(1,y)
    +b_2 G(0,1,1-y,x)+c_1 G(0,0,1,x))+a b_2 c_2 G(0,0,1,y)] + O(\e^4)\\
    %%%%%%%%%%%%%%%%%%%%%
    %%%%%%%% F3 %%%%%%%%%
    F_3(a_1 \e, &a_2 \e, b_1 \e, b_2 \e; 1+ c \e; x , y) =  1 + \e^2 [- a_1 b_1 G(0,1,x)-a_2 b_2 G(0,1,y)] + \e^3 [a_1 b_1 c G(0,0,1,x)\nonumber\\
    &+a_1 b_1 G(0,1,1,x) \left(a_1+b_1-c\right)+a_2 b_2 c G(0,0,1,y)+a_2 b_2 G(0,1,1,y) \left(a_2+b_2-c\right)]\nonumber\\
    &+ O(\e^4)\\
    F_A (a \e,& b_1  \e, b_2 \e, b_3 \e, 1+ c_1 \e, 1+ c_2 \e, 1+ c_3 \e,x,y,z)= 1+ \e^2 [-a b_1 G(0,1,x)-a b_2 G(0,1,y)-a b_3 G(0,1,z)]\nonumber\\
    &+ \e^3 [a b_1 G(0,1,1,x) \left(a+b_1-b_2-b_3-c_1\right)+a b_1 c_1 G(0,0,1,x)+a b_2 G(0,1,1,y) \left(a+b_2-b_3-c_2\right)\nonumber\\
    &+a b_2 c_2 G(0,0,1,y)+a b_3 G(0,1,1,z) \left(a+b_3-c_3\right)+a b_3 c_3 G(0,0,1,z)+a b_1 b_2 G(0,1,x) G(1,y)\nonumber\\
    &+a b_1 b_2 G(0,1,1-y,x)+a b_1 b_3 G(0,1,x) G(1,z)+a b_1 b_3 G(0,1,1-z,x)+a b_3 b_2 G(0,1,y) G(1,z)\nonumber\\
    &+a b_3 b_2 G(0,1,1-z,y)]+ O(\e^4)\\
    F_B(a_1 \e,& a_2 \e, a_3 \e, b_1 \e, b_2 \e, b_3 \e, 1+ c \e; x , y, z) = 1+ \e^2 [a_1 b_1 (-G(0,1,x))-a_2 b_2 G(0,1,y)-a_3 b_3 G(0,1,z)]\nonumber\\
    &+\e^3[a_1 b_1 c G(0,0,1,x)+a_1 b_1 G(0,1,1,x) \left(a_1+b_1-c\right)+a_2 b_2 c G(0,0,1,y)\nonumber\\
    &+a_2 b_2 G(0,1,1,y) \left(a_2+b_2-c\right)+a_3 b_3 c G(0,0,1,z)+a_3 b_3 G(0,1,1,z) \left(a_3+b_3-c\right)]\nonumber\\
    &+ O(\e^4)\\
    F_D(a \e, & b_1 \e, b_2 \e, b_3 \e, 1+ c \e, x ,y ,z)=1+ \e^2 [-a b_1 G(0,1,x)-a b_2 G(0,1,y)-a b_3 G(0,1,z)]\nonumber\\
    &+ \e^3[a b_1 G(0,1,1,x) \left(a+b_1-c\right)-a b_1 \left(b_2+b_3-c\right) G(0,0,1,x)+a b_2 G(0,1,1,y) \left(a+b_2-c\right)\nonumber\\
    &+a b_2 \left(c-b_3\right) G(0,0,1,y)+a b_3 c G(0,0,1,z)+a b_3 G(0,1,1,z) \left(a+b_3-c\right)+a b_1 b_2 G(0,1,x) G(1,y)\nonumber\\
    &-a b_1 b_2 G(1,y) G(0,y,x)+a b_1 b_2 G(0,y,1,x)+a b_1 b_3 G(0,1,x) G(1,z)-a b_1 b_3 G(1,z) G(0,z,x)\nonumber\\
    &+a b_1 b_3 G(0,z,1,x)+a b_2 b_3 G(0,1,y) G(1,z)-a b_2 b_3 G(1,z) G(0,z,y)+a b_2 b_3 G(0,z,1,y)]\nonumber\\
    &+ O(\e^4)\\
    F_S( a_1 \e, &a_2 \e, b_1 \e, b_2 \e, b_3 \e, 1+ c \e, x,y,z)= 1+ \e^2 [a_1 b_1 (-G(0,1,x))-a_2 b_2 G(0,1,y)-a_2 b_3 G(0,1,z)] \nonumber\\
    &+ \e^3[a_1 b_1 c G(0,0,1,x)+a_1 b_1 G(0,1,1,x) \left(a_1+b_1-c\right)+a_2 b_2 G(0,1,1,y) \left(a_2+b_2-c\right)\nonumber\\
    &+a_2 b_2 \left(c-b_3\right) G(0,0,1,y)+a_2 b_3 c G(0,0,1,z)+a_2 b_3 G(0,1,1,z) \left(a_2+b_3-c\right)\nonumber\\
    &+a_2 b_2 b_3 G(0,1,y) G(1,z)-a_2 b_2 b_3 G(1,z) G(0,z,y)+a_2 b_2 b_3 G(0,z,1,y)] + O(\e^4)\\
    F_K(a_1 \e ,&a_2 \e,b_1 \e ,b_2 \e ;1+ c_1 \e,1+ c_2 \e,1+ c_3 \e; x,y,z) = 1+ \e^2 [a_1 b_1 (-G(0,1,x))-a_2 b_2 G(0,1,y)-a_2 b_1 G(0,1,z)] \nonumber\\
    &+ \e^3[a_1 b_1 G(0,1,1,x) \left(a_1-a_2+b_1-c_1\right)+a_1 b_1 c_1 G(0,0,1,x)+a_2 b_2 G(0,1,1,y) \left(a_2-b_1+b_2-c_2\right)\nonumber\\
    &+a_2 b_2 c_2 G(0,0,1,y)+a_2 b_1 G(0,1,1,z) \left(a_2+b_1-c_3\right)+a_2 b_1 c_3 G(0,0,1,z)+a_1 a_2 b_1 G(0,1,x) G(1,z)\nonumber\\
    &+a_1 a_2 b_1 G(0,1,1-z,x)+a_2 b_2 b_1 G(0,1,y) G(1,z)+a_2 b_2 b_1 G(0,1,1-z,y)]+ O(\e^4)\\
    F_M(a_1 \e ,&a_2 \e ,b_1 \e ,b_2 \e ,1+ c_1 \e ,1+ c_2 \e;x,y,z) = 1+ \e^2[a_1 b_1 (-G(0,1,x))-a_2 b_2 G(0,1,y)-a_2 b_1 G(0,1,z)]\nonumber\\
    &+ \e^3[a_1 b_1 G(0,1,1,x) \left(a_1-a_2+b_1-c_1\right)+a_1 b_1 c_1 G(0,0,1,x)+a_2 b_2 G(0,1,1,y) \left(a_2+b_2-c_2\right)\nonumber\\
    &+a_2 b_2 \left(c_2-b_1\right) G(0,0,1,y)+a_2 b_1 G(0,1,1,z) \left(a_2+b_1-c_2\right)+a_2 b_1 c_2 G(0,0,1,z)+a_1 a_2 b_1 G(0,1,x) G(1,z)\nonumber\\
    &+a_1 a_2 b_1 G(0,1,1-z,x)+a_2 b_2 b_1 G(0,1,y) G(1,z)-a_2 b_2 b_1 G(1,z) G(0,z,y)+a_2 b_2 b_1 G(0,z,1,y)]\nonumber\\
    &O(\e^4)\\
    F_N(a_1 \e,&a_2 \e,a_3 \e,b_1 \e,b_2 \e;1+ c_1 \e,1 + c_2 \e;x,y,z) = 1+ \e^2[a_1 b_1 (-G(0,1,x))-a_2 b_2 G(0,1,y)-a_3 b_1 G(0,1,z)]\nonumber\\
    &+\e^3[a_1 b_1 G(0,1,1,x) \left(a_1-a_3+b_1-c_1\right)+a_1 b_1 c_1 G(0,0,1,x)+a_2 b_2 G(0,1,1,y) \left(a_2+b_2-c_2\right)\nonumber\\
    &+a_2 b_2 c_2 G(0,0,1,y)+a_3 b_1 G(0,1,1,z) \left(a_3+b_1-c_2\right)+a_3 b_1 c_2 G(0,0,1,z)+a_1 a_3 b_1 G(0,1,x) G(1,z)\nonumber\\
    &+a_1 a_3 b_1 G(0,1,1-z,x)] + O(\e^4)
\end{align}
\end{scriptsize}

\section{Transformation theory of order two, complete HFs}
\label{sec:Transformation_Theory}

The connection formulae between the Horn's functions are well studied in the literature \cite{Srivastava:1985,Bateman:1953,erdelyi_1948}. We present some of the connection formulae of Appell-Horn functions that are used to compute their series expansions. 
\begin{align}
    &G_1\left(a,b,c,x,y\right)=\left(\frac{1}{x+y+1}\right)^a  \\ \nonumber
    &\quad\times
    F_2 \left(-b-c+1,a,a,1-b,1-c,\frac{-\sqrt{1-4 x y}+2 x+1}{2 (x+y+1)},\frac{-\sqrt{1-4 x y}+2 y+1}{2 (x+y+1)}\right)\\
    %%%
    %%%
    &G_2\left(a,b,c,d,x,y\right)= (x+1)^{-a} (y+1)^{-b}F_2 \left(-c-d+1,a,b,1-c,1-d,\frac{x}{x+1},\frac{y}{y+1}\right)  \label{eqn:G2F2relation}\\
    %%%
    %%%
     &G_3\left(a,b,\frac{x(1-y)}{(1-x)^2},\frac{y(1-x)}{(1-y)^2}\right)=(1-y)^a(1-x)^b G_1\left(a+b,a,b,x,y\right)\\
     %%%
     %%%
    &H_3\left(a,b,c,x,y\right) =(1-4 x)^{-\frac{a}{2}} \left(\frac{-4 x+\sqrt{1-4 x}+1}{2-8 x}\right)^{1-c} \\ \nonumber
    &\quad\times F_3 \left(c-a,a,a-c+1,b,c,\frac{4 x+\sqrt{1-4 x}-1}{8 x-2},\frac{y-\sqrt{1-4 x} y}{2 x}\right)\\
    %%%
    %%%
    &H_4\left(a,b,c,d,x,y\right) = \left(1-\frac{2 \sqrt{x}}{2 \sqrt{x}+1}\right)^a F_2\left(a,c-\frac{1}{2},b,2 c-1,d,\frac{4 \sqrt{x}}{2 \sqrt{x}+1},\frac{y}{2 \sqrt{x}+1}\right)\\
    %%%
    %%%
   &H_6\left(a,b,c,x,y\right)= (4 x+1)^{-\frac{a}{2}} \left(\frac{4 x+\sqrt{4 x+1}+1}{8 x+2}\right)^b \\ \nonumber
    &\hspace{1cm} \times H_2 \left(b,c,a+b,-a-b+1,1-a,-\frac{\left(4 x+\sqrt{4 x+1}+1\right) y}{2 \sqrt{4 x+1}},\frac{-4 x+\sqrt{4 x+1}-1}{8 x+2}\right)\\ 
    %%%
    %%%
    &H_7\left(a,b,c,d,x,y\right) = \left(2 \sqrt{x}+1\right)^{-a} H_2 \left( a,d-\frac{1}{2},b,c,2 d-1,\frac{4 \sqrt{x}}{2 \sqrt{x}+1},2 \sqrt{x} y+y \right)
\end{align}
It is worth mentioning that the correct form of the connection formula between $G_3$ and $G_1$ is found in \cite{Niukkanen_G3_G1}. This expression is brought to the form $G_3\left(a,b;x,y\right)= \dots G_1\left(\dots\right)$ for computation of the series expansion of the former series, which is not presented here due to its long length.

To the best of our knowledge, no connection formula exists for the functions $H_1,H_5$ and $F_4$ having general parameters to other Appell-Horn functions. However, the connection formula of $H_1$ and $F_4$ with some restriction  of parameters can be found in \cite{erdelyi_1948,BrychkovH1} and \cite{Srivastava:1985} respectively, which we present below.

\begin{small}
    \begin{align}
    &H_1\left(d-c,b,c,d,x,y\right)= 2^{-b} \left(-\frac{\sqrt{-4 x y+y^2+2 y+1}+y-1}{(x-1) y}\right)^b\\ \nonumber
    &\quad\times H_2 \left(d-c,b,b,c,d,\frac{-\sqrt{-4 x y+y^2+2 y+1}+y+1}{2 y},\frac{-\sqrt{-4 x y+y^2+2 y+1}-y+1}{2 (x-1)}\right)\\ 
    &H_1\left(a,b,c,\frac{1}{2} (a+b+1),x,y\right)= \left(-2 w-2 \sqrt{(w-1) w}+1\right)^{\frac{b-a}{2}} \left(-2 w+2 \sqrt{(w-1) w}+1\right)^{\frac{1}{2} (-a-b)}\\ \nonumber
    &\quad\times F_2\left(b,\frac{a+b}{2},c,a+b,1-a,\frac{4 \sqrt{(w-1) w}}{2 \sqrt{(w-1) w}+\sqrt{(1-2 w)^2}},-\left(\left(-2 w-2 \sqrt{(w-1) w}+1\right) z\right)\right)
\end{align}
\end{small}

The relations to the Appell $F_4$ are

\begin{align}
    F_4(a,b;c,b,x,y) = \left(1-X\right)^a \left(1-Y\right)^a F_1 \left( a,c-b,a-c+1;c, X ,XY \right)
\end{align}
where,
\begin{align}
    X = \frac{\sqrt{(x+y-1)^2-4 x y}+x+y-1}{2 y}\\
    Y = \frac{\sqrt{(x+y-1)^2-4 x y}+x+y-1}{2 x}
\end{align}

\begin{align}
    F_4(a,b;c,a-c+1,x,y) = F_2 \left( a,b,b;c,a-c+1, X ,Y \right)
\end{align}
where,
\begin{align}
    X &= \frac{1}{2} \left(-\sqrt{(-x+y-1)^2-4 x}+x-y+1\right)\\
    Y &= \frac{1}{2} \left(-\sqrt{(-x+y-1)^2-4 x}-x+y+1\right)
\end{align}
\begin{align}
    F_4(a,b;c,a+b-c+1,x,y) = \, _2F_1\left(a,b;c;X\right) \, _2F_1\left(a,b;a+b-c+1;Y\right)
\end{align}
where,
\begin{align}
   X &= \frac{1}{2} \left(x-y-\sqrt{(-x+y-1)^2-4 x}+1\right)\\
Y &= \frac{1}{2} \left(-x+y-\sqrt{(-x+y-1)^2-4 x}+1\right)
\end{align}

\section{Reduction formulae of MHFs }
\label{sec:Reduction_formulae}

It is possible write a MHF with positive integer values of Pochhammer parameters in terms of simpler functions. These useful reduction formulae have applications in mathematics and physics.
As pointed out in \cite{Hyperdire2}, some reduction formulae of Appell and Lauricella-Saran functions can be immediately derived using the results of their series expansions (Appendix \ref{sec:exactresult}) and the differential reduction formulae from \texttt{HYPERDIRE} \cite{Hyperdire,Hyperdire2} packages. For instance,
\begin{align}
    F_2(1+\e,1+\e&,1+\e;2+\e,2+\e;x,y) = \nonumber\\
    &\Big[\frac{\epsilon +1}{x \epsilon ^2} \theta_x +\frac{\epsilon +1}{y \epsilon ^2} \theta_y +\frac{(\epsilon +1) (x+y-1)}{x y \epsilon ^3 }\theta_y \theta_x\Bigg] \bullet F_2(\e,\e,\e;1+\e,1+\e;x,y)
\end{align}
Using the series expansion of the $F_2$ from Appendix \ref{sec:exactresult}, we find the right-hand side to be
\begin{align}
    F_2(1+\e,1+\e&,1+\e;2+\e,2+\e;x,y) =\nonumber\\
    &\left(\frac{(x+y-1) (G(1;x)-G(1-y;x))}{x y}-\frac{G(1;x)}{x}-\frac{G(1;y)}{y}\right)+O\left(\epsilon ^1\right)
\end{align}

Finally, setting $\e \rightarrow 0$, the reduction formula of Appell $F_2$ reads
\begin{align} \label{eqn:F2reduction}
    F_2(1,1,1;2,2;x,y) &=\frac{(x+y-1) (G(1,x)-G(1-y,x))}{x y}-\frac{G(1,x)}{x}-\frac{G(1,y)}{y}\\
    &= -\frac{(1-x) \log (1-x)}{x y}-\frac{(x+y-1) \log \left(1-\frac{x}{1-y}\right)}{x y}-\frac{\log (1-y)}{y}
\end{align}

The above result differs from the expression obtained in Eq. (A.9) of \cite{F2reduction}. However, we found that \eeqref{eqn:F2reduction} is numerically consistent. 
\begin{comment}

Similarly, the reduction formulae of Appell $F_2$ with other set of positive values of  Pochhammer parameters can be easily obtained by utilizing the \texttt{HYPERDIRE} package and \eeqref{eqn:F2reduction}. As an example
\begin{align}
   F_2(3,2,1;3,2;x,y) = \left(-\frac{1}{x-1}-\frac{1}{x-1} \theta_x-\frac{1}{x-1} \theta_y -\frac{1}{(x-1) x} \theta_y \theta_x\right) \bullet F_2(1,1,1;2,2;x,y)
\end{align}
Thus,
\begin{align} \label{eqn:F2reduction}
    F_2(3,2,1;3,2;x,y) = -\frac{\log (1-x)}{x^2 y}+\frac{\log \left(\frac{x+y-1}{y-1}\right)}{x^2 y}+\frac{1}{(x-1) x (x+y-1)}
\end{align}

which matches with the result from literature \cite{F2reduction}.

\end{comment}

We provide a short list of reduction formulae of Appell $F_1, F_3, F_4$, Lauricella - Saran $F_D$ and $F_S$ functions below in terms of MPLs. 

\begin{align}
    F_1(1,1,1;2;x,y) &= \frac{G(1;y)-G(1;x)}{x-y}\\
    F_3(1,1,1,1;2;x,y) &=\frac{G(1;x)+G(1;y)}{x y-x-y}\\
    F_4(1,1;1,1;x,y)&= \frac{1}{\sqrt{x^2-2 x y-2 x+y^2-2 y+1}}\\
    F_D(1,1,1,1;2;x,y,z) &=-\frac{x G(1;x)}{(x-y) (x-z)}+\frac{y G(1;y)}{(x-y) (y-z)}+\frac{z G(1;z)}{(x-z) (z-y)}\\
    F_S(1,1,1,1,1;2;x,y,z)&=-\frac{x G(1;x)}{(x (y-1)-y) (x (z-1)-z)}+\frac{y G(1;y)}{(x (y-1)-y) (y-z)}\nonumber\\
    &-\frac{z G(1;z)}{(x (z-1)-z) (y-z)}\label{eqn:FSreduction}
\end{align}

\begin{comment}
\begin{align}
    F_1(1,1,1;2;x,y) &= \frac{\log (1-y)-\log (1-x)}{x-y}\\
    F_3(1,1,1,1;2;x,y) &= -\frac{\log (1-x)}{- x y +x+y}-\frac{\log (1-y)}{-x y+x+y}\\
    F_4(1,1;1,1;x,y)&= \frac{1}{\sqrt{x^2-2 x y-2 x+y^2-2 y+1}}\\
    F_D(1,1,1,1;2;x,y,z) &=-\frac{x \log (1-x)}{(x-y) (x-z)}+\frac{y \log (1-y)}{(x-y) (y-z)}+\frac{z \log (1-z)}{(x-z) (z-y)}\\
    F_S(1,1,1,1,1;2;x,y,z)&=-\frac{x \log (1-x)}{(x (y-1)-y) (x (z-1)-z)}+\frac{y \log (1-y)}{(x (y-1)-y) (y-z)}\nonumber\\
    &-\frac{z \log (1-z)}{(x (z-1)-z) (y-z)}\label{eqn:FSreduction}
\end{align}    
\end{comment}

These expressions can be written in terms of ordinary logarithms, since
\begin{align*}
    G(1;x) = -\text{Li}_1(x) = \log(1-x)
\end{align*}

Making good use of the \texttt{HYPERDIRE} packages and the expressions from Appendix \ref{sec:exactresult}, a huge number of reduction formulae can be easily derived. This process of obtaining the reduction formula is encoded in the command \texttt{ReduceFunction} command of the presented package.

%The author would like to thank B. Ananthanarayan and Thomas Gehrmann  

\bibliographystyle{JHEP}
%\bibliography{UZh.bib,ref_eps_expansion.bib,refGKZ.bib,refOlsson.bib}
\bibliography{UZH,ref_eps_expansion,refGKZ,refOlsson,refMOBMB}

\end{document}